\documentclass[sigconf,nonacm]{acmart}

\usepackage{booktabs}
\usepackage{multirow}
\usepackage{subcaption}
\usepackage{xspace}
\usepackage{enumitem}
\usepackage{amsmath}

\graphicspath{{./images/}}

\setcopyright{none}
\acmConference[Conference acronym 'XX]{Make sure to enter the correct
  conference title from your rights confirmation email}{June 03--05,
  2018}{Woodstock, NY}

\begin{document}

\title{Beyond Positive Signals: Unlocking Implicit Negative Behaviors for Enhanced Sequential User Modeling}

\author{Zexuan Cheng, Yue Liu, Jun Zhang, Jie Jiang}
\affiliation{\institution{Tencent Inc., Beijing, China}\country{}}
\email{{zexuancheng, herculesliu, neoxzhang, zeus}@tencent.com}

\begin{abstract}
User behavior sequence modeling has become a central component in modern click-through rate (CTR) prediction. Over the past years, the community has invested substantial effort into improving \emph{how} sequences are encoded, from target-aware attention and interest evolution networks to unified architectures that jointly process sequential and non-sequential features. However, a more fundamental question remains under-explored: \emph{what} should constitute the behavior sequence? Current practice constructs sequences exclusively from positive interactions (clicks, purchases, completions), while the far more abundant implicit negative behaviors (skips, low engagement, scroll-past) are largely underutilized. As gains from longer positive sequences approach diminishing returns, we revisit this underutilized data source within the sequential modeling framework. In this paper, we demonstrate that mixed-polarity behavior sequences, which chronologically interleave positive and negative tokens within a fixed length budget, consistently outperform positive-only sequences across diverse model architectures with negligible additional computational overhead. We further identify a semantic indistinguishability problem inherent to naive polarity embeddings and propose \textbf{Target-Aware Polarity Fusion (TAPF)}, a lightweight target-conditioned gating mechanism that provides additional gains by differentiating behavioral evidence. Notably, even the simpler polarity bias baseline captures the majority of improvement, underscoring that the primary contribution is the mixed-polarity data paradigm itself. Comprehensive analyses reveal that mixed-polarity models spontaneously learn polarity-discriminative attention, exhibiting markedly different attention allocation patterns across samples with positive versus negative labels, and deliver gains in both informationally sparse scenarios (low-activity users, cold-start items) and data-rich settings. Experiments on three public benchmarks demonstrate consistent improvements of +1.9\% to +9.6\% relative AUC across five architectures, which validate the practical value of our approach.
\end{abstract}



\maketitle

\begin{figure}[t]
\centering
\includegraphics[width=\columnwidth]{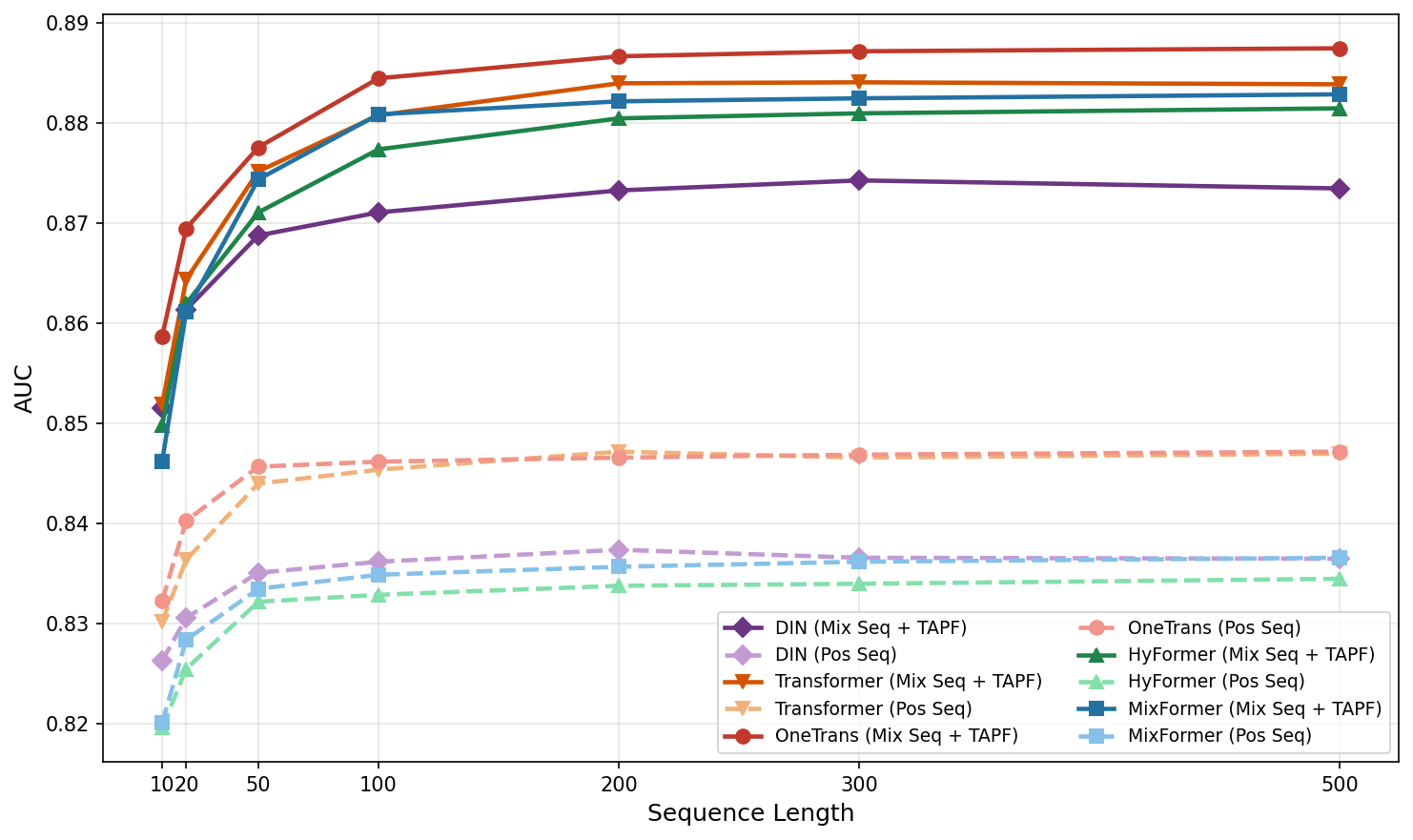}
\caption{Sequence length scaling on KuaiRec across five architectures. Solid lines: mixed-polarity sequences (Mix Seq + TAPF, $r{=}0.5$); dashed lines: positive-only sequences. Mixed sequences achieve substantially higher AUC and exhibit steeper scaling slopes across all models.}
\label{fig:teaser}
\end{figure}

\section{Introduction}
\label{sec:intro}

Click-through rate (CTR) prediction drives modern recommender systems and advertising platforms, determining which content reaches users and governing platform revenue~\cite{guo2017deepfm,zhou2018din}. Sequential behavior modeling, which encodes a user's chronologically ordered interaction history into a dense representation, has emerged as one of the most impactful paradigms for this task~\cite{zhou2019dien,pi2020search,chang2023twin}.

The field has witnessed remarkable progress in \emph{how} to model behavior sequences. Target-aware attention in DIN~\cite{zhou2018din} enabled extracting candidate-relevant signals from user histories. DIEN~\cite{zhou2019dien} captured interest evolution, MIMN~\cite{pi2019practice} and SIM~\cite{pi2020search} addressed ultra-long sequences, and recently, unified architectures such as OneTrans~\cite{onetrans2025}, HyFormer~\cite{hyformer2025}, and MixFormer~\cite{mixformer2025} have integrated sequence modeling and feature interaction within a single backbone. Yet across this entire evolution, an equally important question remains comparatively under-explored: \emph{what} should go into the behavior sequence?

Today, the answer is taken for granted: behavior sequences consist exclusively of \emph{positive} interactions, i.e., clicks, purchases, and completions. The vastly more numerous \emph{implicit} negative behaviors, such as skips, rapid scroll-past, and low watch-completion, are largely underutilized. Unlike \emph{explicit} negative feedback (e.g., user-initiated dislikes or thumbs-down ratings), which is rare and requires deliberate action, implicit negatives arise naturally and abundantly from every user session, reflecting items that were exposed but failed to elicit engagement. We argue that this design choice of excluding them, while historically understandable, leaves substantial value unrealized.

\textbf{Why has negative behavioral context been overlooked in sequential modeling?} While implicit negative behaviors have been explored through multi-channel feedback separation, auxiliary training objectives, or domain-specific feedback taxonomies, they remain largely absent from mainstream behavior sequence modeling. We identify three converging factors. First, the research community has largely adopted a \emph{preference-centric view} of user modeling: positive behaviors are seen as clear indicators of user interest, while implicit negatives are perceived as ambiguous and offering low return on modeling investment. Second, the clear and measurable gains from extending positive sequences (longer histories, better attention mechanisms, more efficient retrieval) have absorbed the field's attention, leaving negative behaviors unexplored within the sequential modeling paradigm. Third, incorporating negatives into the same sequence creates a \emph{semantic conflict}: positive and negative tokens occupy the same embedding space but carry opposing semantics. Prior attempts to leverage negative feedback have therefore resorted to dual-stream architectures or separate encoders~\cite{chen2020bias}, isolating negatives from the main sequence modeling pipeline.

\textbf{Why revisit this now?} The motivation is both practical and principled. On the practical side, as sequence modeling capabilities have been continuously enhanced, the gains from extending positive-only sequences are approaching \emph{diminishing returns}: positive behavior histories are bounded by the user's actual engagement rate(particularly in advertising domains where positive interactions are extremely sparse), while implicit negative histories are considerably longer (e.g., the positive behavior rate is 7.4\% on KuaiRec, 37.9\% on KuaiRand, and 6.3\% on TAAC). On the principled side, a user representation built solely from preferences may be incomplete: it captures what a user likes but not what they reject, limiting the model's ability to discriminate between superficially similar items that receive very different user responses.

This confluence raises a natural question: \emph{Can implicit negative behaviors provide effective information for CTR prediction?} In this paper, we answer affirmatively. As shown in Figure~\ref{fig:teaser}, incorporating implicit negative behaviors into behavior sequences yields substantial improvements: on KuaiRec, mixed-polarity sequences outperform positive-only sequences by 4.2\%--5.5\% relative AUC improvement across architectures, with the gap widening as the sequence length budget increases.

Concretely, this paper makes three contributions:
\textbf{(1)} We demonstrate that mixed-polarity sequences, which chronologically interleave positive and negative tokens within the same length budget, consistently outperform positive-only sequences across both traditional two-stage and unified-block architectures, with negligible additional computational overhead. We emphasize that this is primarily a data-level contribution: under identical computational budgets, simply replacing a portion of positive tokens with chronologically recent negative tokens yields the majority of observed gains, suggesting that the community may be under-utilizing readily available behavioral signals. We hope this finding motivates future research to investigate mixed-polarity behavior sequences beyond the conventional focus on positive-only sequences.
\textbf{(2)} We identify the semantic indistinguishability problem that arises when positive and negative tokens share an embedding space, and propose \textbf{Target-Aware Polarity Fusion (TAPF)}, a lightweight gating mechanism that enables target-conditioned differentiation of behavioral evidence. Experiments show that TAPF achieves competitive performance across diverse architectures and datasets, serving as a good starting baseline for future mixed-polarity sequence modeling.
\textbf{(3)} We provide comprehensive mechanistic analyses, including attention polarity gap, user activity stratification, cold-start evaluation, and negative behavior ablation, explaining \emph{why} and \emph{when} negative behavioral context is most effective.

\section{Related Work}
\label{sec:related}

\subsection{Traditional Recommendation Paradigms}

Industrial CTR prediction has long followed a \emph{two-stage pipeline} in which a feature interaction module and a sequence modeling module operate independently, with their outputs concatenated before a final prediction head.

\textbf{Feature Interaction.} Early approaches such as logistic regression~\cite{richardson2007predicting}, factorization machines~\cite{rendle2010fm}, and Wide \& Deep~\cite{cheng2016wide} established the paradigm of learning feature interactions for CTR prediction. DeepFM~\cite{guo2017deepfm} automated second-order crossing via an FM layer, and DCN~\cite{wang2017deep}/DCN-V2~\cite{wang2021dcnv2} introduced explicit cross layers for higher-order interactions. More recently, the field has shifted toward architectures that treat feature interaction as a token-mixing or attention problem at scale. Hiformer~\cite{hiformer2023} proposed heterogeneous self-attention that accounts for the diverse nature of recommendation features. Wukong~\cite{wukong2024} established the first empirical scaling law for recommendation using stacked factorization machine blocks, demonstrating consistent quality gains with increased compute. RankMixer~\cite{rankmixer2024} replaced self-attention with hardware-efficient multi-head token mixing, improving GPU utilization from 4.5\% to 45\% while scaling to 1B parameters. TokenMixer-Large~\cite{tokenmixerlarge2026} further extended this approach to 7--15B parameters via sparse token-level mixture-of-experts. UniMixer~\cite{unimixer2024} unified attention-based, token-mixing, and FM-based methods into a single learnable architecture with validated scaling behavior. Despite their effectiveness at capturing cross-feature patterns, all these methods treat each prediction instance as a static feature vector without modeling temporal user behavior dynamics.

\textbf{Sequence Modeling.} DIN~\cite{zhou2018din} introduced target-aware attention pooling for user behavior sequences, and DIEN~\cite{zhou2019dien} further modeled interest evolution. Transformer-based approaches such as BST~\cite{chen2019behavior}, SASRec~\cite{kang2018sasrec}, and BERT4Rec~\cite{sun2019bert4rec} demonstrated the effectiveness of self-attention for sequential patterns. As user histories grew to thousands of interactions, scalability became the central challenge. MIMN~\cite{pi2019practice} employed external memory, SIM~\cite{pi2020search} proposed two-stage retrieval-then-attention, ETA~\cite{eta2022} introduced end-to-end SimHash-based target attention for sub-linear sequence search, and TWIN~\cite{chang2023twin} scaled target attention to full lifetime histories via cascading retrieval. Most recently, LONGER~\cite{longer2024} addressed ultra-long sequences (10,000+ tokens) through a Global Token mechanism and Token Merge module that maintains full temporal context while reducing quadratic complexity, enabling production deployment across multiple scenarios at ByteDance.

While architecturally diverse, these traditional approaches share a structural limitation: the feature interaction and sequence modeling modules remain \emph{decoupled}, with limited cross-modal information flow during encoding.

\subsection{Unified Recommendation Architectures}

Recognizing the inefficiency of decoupled pipelines, a recent line of work proposes \emph{unified architectures} that jointly model feature interactions and sequential behavior within a single backbone. The core insight is that treating behavior tokens and feature tokens as inputs to a shared attention mechanism enables richer cross-modal interactions during encoding rather than only at the late-fusion stage.

OneTrans~\cite{onetrans2025} formalized this idea by processing candidate feature tokens and historical behavior tokens within one Transformer block, using causal masking for autoregressive interest modeling. HyFormer~\cite{hyformer2025} extended the unified paradigm to long sequences through Query Decoding (expanding non-sequential features into global tokens for decoding behavior sequences) and Query Boosting (cross-query heterogeneous interaction), eliminating the separate compression stage. MixFormer~\cite{mixformer2025} employs token-mixing layers that alternate between sequence-level and feature-level mixing, jointly scaling dense feature interaction capacity and sequence length within a single backbone. MTGR~\cite{mtgr2025} preserves traditional cross features within a generative recommendation framework, introducing user-level sequence compression for deployment efficiency.

These unified architectures represent the current state of the art. However, they share a common characteristic with traditional models: their behavior sequences contain \emph{exclusively positive interactions}, leaving the potential signal in implicit negative behaviors largely unexploited.

\subsection{Negative Behavior Modeling}

An important distinction exists between \emph{explicit} negative feedback (e.g., user-initiated dislikes, thumbs-down ratings) and \emph{implicit} negative behaviors (e.g., skips, low watch-completion, rapid scroll-past without engagement). Explicit negatives are relatively scarce and require deliberate user action, whereas implicit negatives are considerably more abundant, arising naturally within every user session, and exhibit higher semantic ambiguity since the absence of engagement may reflect disinterest, inattention, or contextual factors. These differences in \emph{prevalence} and \emph{modeling complexity} make the two settings substantially different in practice.

The majority of existing work on negative behavior modeling addresses \emph{explicit} negatives or employs negatives solely as supervisory signals during training. In collaborative filtering, BPR~\cite{rendle2009bpr} and hard negative mining strategies~\cite{zhang2013optimizing,ding2020simplify} leverage negatives to formulate pairwise ranking objectives but do not encode them into user representations. Contrastive methods such as CL4Rec~\cite{xie2022contrastive} construct synthetic negative views for self-supervised pre-training, and denoising approaches~\cite{wang2021denoising} filter noisy positives without modeling the underlying negative behavioral patterns.

Among efforts that incorporate negative feedback into representation learning, DFN~\cite{xue2019dfn} models explicit dislikes through a dedicated feedback interaction module, treating negatives as a separate information channel. XDM~\cite{ouyang2019xdm} leverages unclicked items in matching-stage retrieval via set-level contrastive signals. FeedRec~\cite{wu2022feedrec} encodes heterogeneous feedback types (clicks, skips, finish reads) through feedback-type-specific interaction modules for news recommendation, and introduces an additional disentangling loss to separate representations of different feedback types. These methods primarily rely on \emph{explicit} negative signals or domain-specific feedback taxonomies. Since explicit negatives are sparse and qualitatively different from the large volume of implicit negatives, and our setting involves sequences containing hundreds of semantically ambiguous implicit negative behaviors per user, these architectures are difficult to adapt to our problem and are thus not included as baselines in our experiments.

In this work, we treat \emph{implicit} negative behaviors as \emph{first-class sequence tokens} that participate in the same chronological attention computation as positive tokens. Rather than confining negatives to auxiliary modules or relying on explicit feedback signals, we integrate the abundant implicit negatives into the user's unified behavioral representation, augmented with polarity-aware mechanisms for semantic calibration.

\section{Methodology}
\label{sec:method}

\subsection{Overview}

\begin{figure*}[t]
\centering
\includegraphics[width=0.85\textwidth]{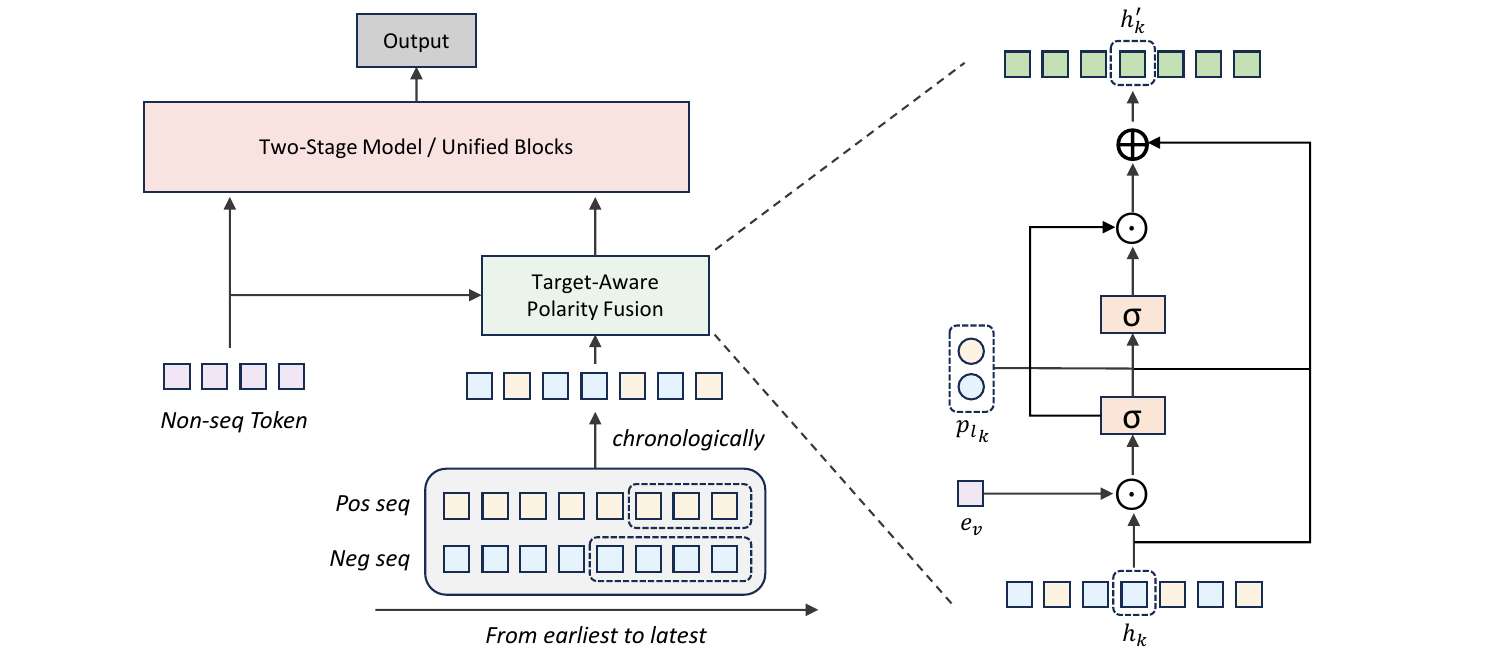}
\caption{Framework overview. \textbf{Left}: Positive and negative behavior sequences are merged chronologically into a mixed-polarity sequence. \textbf{Center}: The TAPF module processes the mixed sequence through four stages: polarity bias embedding, sequence encoding, target-conditioned gating, and weighted aggregation. \textbf{Right}: The output user representation is combined with non-sequence features and fed to any CTR model (two-stage or unified-block).}
\label{fig:framework}
\end{figure*}

Figure~\ref{fig:framework} presents the overall framework. Given a user's positive and negative behavior histories, we construct a chronologically interleaved mixed-polarity sequence (\S\ref{sec:construction}), encode it with polarity-aware semantic calibration (\S\ref{sec:polarity_calibration}), and apply target-conditioned aggregation via TAPF (\S\ref{sec:tapf}). The resulting user representation is compatible with any downstream CTR architecture.

\subsection{Problem Formulation}

We consider the standard CTR prediction setting. Let $\mathcal{U}$ denote the set of users and $\mathcal{I}$ the set of items. Each user $u \in \mathcal{U}$ is associated with a set of contextual features $\mathbf{x}_u$ (e.g., demographics, device type) and an ordered interaction history with the platform. We decompose this history based on behavioral polarity:
\begin{itemize}[leftmargin=*,topsep=2pt]
\item \textbf{Positive history} $\mathcal{H}^+_u = \{(i^+_1, t^+_1), (i^+_2, t^+_2), \ldots, (i^+_{N^+}, t^+_{N^+})\}$: items with which the user engaged positively (click, purchase, high watch-completion), ordered by timestamp.
\item \textbf{Implicit negative history} $\mathcal{H}^-_u = \{(i^-_1, t^-_1), \ldots, (i^-_{N^-}, t^-_{N^-})\}$: items exposed to the user but met with disengagement (skip, low watch-completion, rapid scroll-past).
\end{itemize}

Given a candidate item $v \in \mathcal{I}$ with features $\mathbf{x}_v$, existing sequential CTR models construct a behavior sequence solely from $\mathcal{H}^+_u$ to represent the user. Our framework instead constructs a \emph{mixed-polarity sequence} $\mathcal{S}_u$ by drawing from both $\mathcal{H}^+_u$ and $\mathcal{H}^-_u$, and the CTR prediction task becomes estimating:
\begin{equation}
\hat{y} = f(\mathcal{S}_u, \mathbf{x}_u, \mathbf{x}_v) = P(y = 1 \mid \mathcal{S}_u, \mathbf{x}_u, \mathbf{x}_v)
\end{equation}
where $\mathcal{S}_u$ encodes both positive and negative behavioral signals. We introduce polarity-aware encoding mechanisms to enable effective joint modeling of the mixed sequence.

\subsection{Chronologically-Interleaved Mix-Polarity Sequence Construction}
\label{sec:construction}

Given a fixed sequence length budget $L$ and a positive ratio hyperparameter $r \in [0,1]$, we construct the mixed-polarity sequence as follows:
\begin{enumerate}[leftmargin=*,topsep=2pt]
\item Sample the most recent $\lfloor rL \rfloor$ interactions from $\mathcal{H}^+_u$ and the most recent $\lfloor (1{-}r)L \rfloor$ interactions from $\mathcal{H}^-_u$.
\item Merge by timestamp into a single chronologically ordered sequence $\mathcal{S}_u = \{(i_k, t_k, p_k)\}_{k=1}^L$, where $p_k \in \{+1,-1\}$ denotes the polarity of each token.
\end{enumerate}
Chronological ordering preserves the natural temporal context of user behavior (e.g., a skip followed by a click on a related item). Importantly, the total sequence length $L$ remains unchanged: negative tokens \emph{replace} a portion of positive tokens rather than appending to them, keeping the inference cost of the sequence encoder identical.

\subsection{Polarity-Aware Semantic Calibration}
\label{sec:polarity_calibration}

Once the mixed-polarity sequence is constructed, a fundamental semantic challenge arises: positive and negative tokens share the same item embedding space, making them indistinguishable to the encoder without explicit polarity information. We address this through a progression from a naive baseline to our proposed solution.

\subsubsection{Naive Approach: Polarity Bias Embedding (PBE).}

The most straightforward approach is to augment each token with a learnable polarity-type embedding, analogous to segment embeddings in language models~\cite{devlin2019bert}. The initial representation $\mathbf{h}^{(0)}_k$ for the $k$-th token in the sequence is computed as:

\begin{equation}
\mathbf{h}^{(0)}_k = \mathbf{e}_{i_k} + \mathbf{b}_{p_k} + \mathbf{pos}_k
\end{equation}
where $\mathbf{e}_{i_k} \in \mathbb{R}^d$ is the item embedding, $\mathbf{b}_{p_k} \in \mathbb{R}^d$ is a learnable polarity embedding (zero-initialized for stable optimization), and $\mathbf{pos}_k$ is the positional encoding. This provides the encoder with an explicit signal to distinguish positive from negative tokens.

\subsubsection{Limitation of PBE}

\begin{figure}[t]
\centering
\includegraphics[width=\columnwidth]{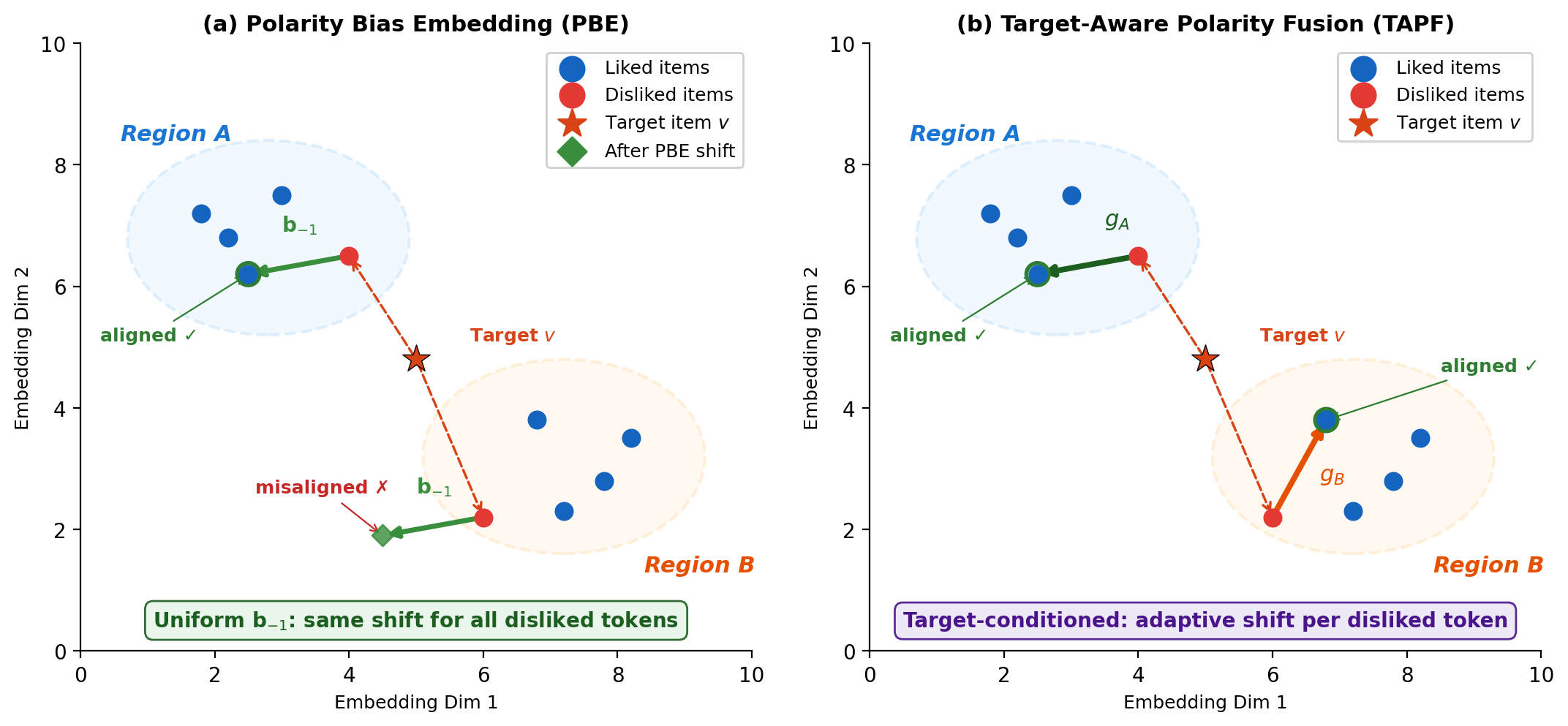}
\caption{Comparison of PBE and TAPF. (a)~PBE applies a uniform polarity bias to negative tokens across different interest regions, making it difficult to accurately represent them within the user's semantic interest space. (b)~TAPF introduces target item information to condition the calibration per token, enabling the model to adaptively adjust each negative token's representation according to its relevance to the candidate, thereby achieving interest-region-aware polarity encoding.}
\label{fig:pbe_tapf}
\end{figure}

While PBE resolves basic type disambiguation, it applies a \emph{uniform, item-agnostic} offset to all negative tokens: every negative interaction receives the identical $\mathbf{b}_{-1}$ regardless of which specific item was skipped or how semantically related that item is to the current candidate. As illustrated in Figure~\ref{fig:pbe_tapf}(a), this rigidity prevents the model from distinguishing between a highly informative negative signal (e.g., skipping an item in the same category as the candidate) and an uninformative one (e.g., skipping an entirely unrelated item). The semantic role of a negative behavior is inherently \emph{relative to the target item}, and a static bias cannot capture this relativity.

\subsubsection{Our Solution: Target-Aware Polarity Fusion (TAPF).}
\label{sec:tapf}

To resolve the semantic indistinguishability of PBE, we propose \textbf{Target-Aware Polarity Fusion (TAPF)}, which introduces target-conditioned, position-wise gated interaction to achieve polarity-aware semantic calibration. The computation proceeds in three steps:

\textbf{Step 1: Target-Aware Interaction.} For each sequence token $\mathbf{h}_k$, we compute its element-wise interaction with the target item embedding $\mathbf{e}_v$, then project through a non-linear layer to obtain the interaction feature:
\begin{equation}
\mathbf{e}_k = \text{ReLU}(\mathbf{W}_{\text{int}}(\mathbf{h}_k \odot \mathbf{e}_v))
\end{equation}
where $\mathbf{W}_{\text{int}} \in \mathbb{R}^{d \times d}$ is a learnable projection and $\odot$ denotes element-wise multiplication. This captures the relevance between each historical behavior and the current candidate.

\textbf{Step 2: Content-Adaptive Gating.} A gating vector is computed from the concatenation of the original token, the interaction feature, and a polarity embedding:
\begin{equation}
\mathbf{g}_k = \sigma(\mathbf{W}_g[\mathbf{h}_k \,\|\, \mathbf{e}_k \,\|\, \mathbf{b}_{p_k}]) \in (0,1)^d
\end{equation}
where $\mathbf{b}_{p_k} \in \mathbb{R}^{d_p}$ is a learnable polarity embedding indexed by $p_k \in \{+1,-1\}$ as defined in \S\ref{sec:construction}, and $\mathbf{W}_g \in \mathbb{R}^{d \times (2d + d_p)}$. The gate operates at full dimensionality $d$, enabling fine-grained per-dimension control.

\textbf{Step 3: Signed Residual Fusion.} The calibrated representation is computed via a signed residual using the polarity sign $p_k$ directly:
\begin{equation}
\mathbf{h}'_k = \mathbf{h}_k + p_k \cdot (2\mathbf{g}_k - 1) \odot \mathbf{e}_k
\end{equation}
The term $(2\mathbf{g}_k - 1) \in (-1, +1)^d$ acts as a learned scaling factor that, combined with the polarity sign $p_k$, enables the model to adaptively push positive tokens toward the target while pushing negative tokens away, with the magnitude and direction controlled independently per dimension.

\textbf{Computational Cost.} TAPF consists solely of linear projections and element-wise operations, introducing linear computational complexity that is negligible relative to the backbone encoder's cost.

\section{Experiments}
\label{sec:exp}

We design experiments around four research questions:
\begin{itemize}[leftmargin=*,topsep=2pt]
\item \textbf{RQ1 (Cross-Architecture Generality)}: For architecturally diverse models, including traditional two-stage pipelines and unified-block frameworks, does replacing positive-only behavior sequences with mixed-polarity sequences yield consistent improvements? Is the benefit specific to certain architectures, or does it generalize as a model-agnostic principle?
\item \textbf{RQ2 (Sequence Length Scaling)}: As the behavior sequence grows longer, does the mixed-polarity sequence exhibit expected performance gains? Compared to positive-only sequences under the same length budget, does it achieve superior scaling efficiency, i.e., more performance per additional behavior token?
\item \textbf{RQ3 (Composition Ratio)}: How does the ratio between positive and negative behaviors within the mixed sequence affect model performance? Is there a universally optimal composition, and how sensitive is the model to this hyperparameter?
\item \textbf{RQ4 (Mechanistic Understanding)}: Why do mixed-polarity sequences improve CTR prediction? What internal mechanisms do they activate? For positive and negative test samples, and for items of different popularity levels, are the gains uniform or concentrated in specific subgroups?
\end{itemize}

\subsection{Experimental Setup}

\subsubsection{Datasets.}
We evaluate on three public benchmarks summarized in Table~\ref{tab:datasets}. \textbf{KuaiRec}~\cite{gao2022kuairec} is a densely observed short-video dataset from Kuaishou; positive labels are defined as watch ratio $\geq$2.0. \textbf{KuaiRand}~\cite{gao2022kuairand} is a randomly exposed recommendation log from Kuaishou with unbiased labels; its random-exposure protocol yields noisier positives, serving as a robustness stress test. \textbf{TAAC-2025} is an industrial-scale advertising CTR dataset from Tencent Ads with click/conversion labels. KuaiRec is split by users; KuaiRand and TAAC are split temporally.

\begin{table}[t]
\centering
\caption{Dataset statistics. Pos.\ Rate: fraction of positive interactions. Max Pos.\ Len.\: maximum positive behavior sequence length per user. Max Full Len.\: maximum full (positive + negative) behavior sequence length per user.}
\label{tab:datasets}
\small
\begin{tabular}{@{}l|rrr@{}}
\toprule
& \textbf{KuaiRec} & \textbf{KuaiRand} & \textbf{TAAC-2025} \\
\midrule
\#Users & 7,176 & 1,000 & 9.4M \\
\#Items & 10,728 & 4.37M & 2.3M \\
\#Interactions & 12.5M & 5.1M & 174M \\
Pos.\ Rate & 7.4\% & 37.9\% & 6.3\% \\
Max Pos.\ Len.\ & 2,244 & 20,487 & 57 \\
Max Full Len.\ & 16,015 & 127,647 & 99 \\
\bottomrule
\end{tabular}
\end{table}

\subsubsection{Baseline Architectures.}
We evaluate five models spanning two paradigms. (1)~\textbf{Traditional two-stage}: \textbf{DIN}~\cite{zhou2018din} and \textbf{Transformer} (a standard Transformer encoder for sequence modeling). (2)~\textbf{Unified-block}: \textbf{OneTrans}~\cite{onetrans2025}, \textbf{HyFormer}~\cite{hyformer2025}, and \textbf{MixFormer}~\cite{mixformer2025}. For each model, we evaluate three sequence settings: \textbf{Pos Seq} (positive-only, $r{=}1.0$), \textbf{Mix Seq + PB} (mixed-polarity with Polarity Bias Embedding, $r{=}0.5$), and \textbf{Mix Seq + TAPF} (mixed-polarity with TAPF, $r{=}0.5$).

\subsubsection{Implementation Details.}
All models use embedding dimension $d{=}64$, batch size 2048, Adam optimizer (lr $10^{-3}$; $5{\times}10^{-4}$ for KuaiRand), and are trained for 1 epoch. Sequence length $L{=}100$ for KuaiRec/KuaiRand and $L{=}50$ for TAAC. The positive ratio is $r{=}0.5$ for the main experiment; full ratio sweeps are in RQ3. We report AUC as the primary metric, averaged over 5 random seeds with standard deviations shown as subscripts.

\subsection{RQ1: Cross-Architecture Generality}

We begin by examining whether mixed-polarity sequences provide consistent improvements across architecturally diverse models, a critical question because a truly model-agnostic technique should benefit any backbone.Table~\ref{tab:main} presents the main results. For each architecture, we report the positive-only baseline  alongside mixed-polarity variants with PB  and TAPF.

\begin{table*}[t]
\centering
\caption{Main experimental results (5-seed mean). Standard deviations shown as gray subscripts. $\Delta$: absolute AUC improvement over Pos Seq. Imp.: relative improvement (\%).}
\label{tab:main}
\small
\begin{tabular}{@{}lll|ccc|ccc|ccc@{}}
\toprule
& & & \multicolumn{3}{c|}{\textbf{KuaiRec}} & \multicolumn{3}{c|}{\textbf{KuaiRand}} & \multicolumn{3}{c}{\textbf{TAAC-2025}} \\
\textbf{Type} & \textbf{Model} & \textbf{Sequence} & AUC & $\Delta$ & Imp. & AUC & $\Delta$ & Imp. & AUC & $\Delta$ & Imp. \\
\midrule
\multirow{8}{*}{Two-Stage}
& \multirow{4}{*}{DIN}
& Pos Seq & .8362{\color{gray}\scriptsize$_{\pm29}$} & --- & --- & .6094{\color{gray}\scriptsize$_{\pm35}$} & --- & --- & .7437{\color{gray}\scriptsize$_{\pm23}$} & --- & --- \\
& & Pos Seq + TAPF & .8387{\color{gray}\scriptsize$_{\pm24}$} & +.0025 & 0.3\% & .6098{\color{gray}\scriptsize$_{\pm31}$} & +.0004 & 0.1\% & .7432{\color{gray}\scriptsize$_{\pm26}$} & $-$.0005 & $-$0.1\% \\
& & Mix Seq + PB & .8650{\color{gray}\scriptsize$_{\pm26}$} & +.0288 & 3.4\% & .6624{\color{gray}\scriptsize$_{\pm31}$} & +.0530 & 8.7\% & .7557{\color{gray}\scriptsize$_{\pm15}$} & +.0120 & 1.6\% \\
& & Mix Seq + TAPF & \textbf{.8711}{\color{gray}\scriptsize$_{\pm25}$} & \textbf{+.0349} & \textbf{4.2\%} & \textbf{.6655}{\color{gray}\scriptsize$_{\pm29}$} & \textbf{+.0561} & \textbf{9.2\%} & \textbf{.7607}{\color{gray}\scriptsize$_{\pm22}$} & \textbf{+.0170} & \textbf{2.3\%} \\
\cmidrule{2-12}
& \multirow{4}{*}{Transformer}
& Pos Seq & .8454{\color{gray}\scriptsize$_{\pm32}$} & --- & --- & .6113{\color{gray}\scriptsize$_{\pm27}$} & --- & --- & .7390{\color{gray}\scriptsize$_{\pm21}$} & --- & --- \\
& & Pos Seq + TAPF & .8488{\color{gray}\scriptsize$_{\pm24}$} & +.0034 & 0.4\% & .6126{\color{gray}\scriptsize$_{\pm26}$} & +.0013 & 0.2\% & .7422{\color{gray}\scriptsize$_{\pm25}$} & +.0032 & 0.4\% \\
& & Mix Seq + PB & .8719{\color{gray}\scriptsize$_{\pm28}$} & +.0265 & 3.1\% & .6654{\color{gray}\scriptsize$_{\pm30}$} & +.0541 & 8.8\% & .7514{\color{gray}\scriptsize$_{\pm26}$} & +.0124 & 1.7\% \\
& & Mix Seq + TAPF & \textbf{.8808}{\color{gray}\scriptsize$_{\pm24}$} & \textbf{+.0354} & \textbf{4.2\%} & \textbf{.6698}{\color{gray}\scriptsize$_{\pm28}$} & \textbf{+.0585} & \textbf{9.6\%} & \textbf{.7570}{\color{gray}\scriptsize$_{\pm21}$} & \textbf{+.0180} & \textbf{2.4\%} \\
\midrule
\multirow{12}{*}[-0.5em]{Unified Block}
& \multirow{4}{*}{OneTrans}
& Pos Seq & .8462{\color{gray}\scriptsize$_{\pm24}$} & --- & --- & .6209{\color{gray}\scriptsize$_{\pm25}$} & --- & --- & .7349{\color{gray}\scriptsize$_{\pm22}$} & --- & --- \\
& & Pos Seq + TAPF & .8497{\color{gray}\scriptsize$_{\pm25}$} & +.0035 & 0.4\% & .6228{\color{gray}\scriptsize$_{\pm30}$} & +.0019 & 0.3\% & .7358{\color{gray}\scriptsize$_{\pm28}$} & +.0009 & 0.1\% \\
& & Mix Seq + PB & .8777{\color{gray}\scriptsize$_{\pm27}$} & +.0315 & 3.7\% & .6602{\color{gray}\scriptsize$_{\pm22}$} & +.0393 & 6.3\% & .7428{\color{gray}\scriptsize$_{\pm23}$} & +.0079 & 1.1\% \\
& & Mix Seq + TAPF & \textbf{.8845}{\color{gray}\scriptsize$_{\pm20}$} & \textbf{+.0383} & \textbf{4.5\%} & \textbf{.6623}{\color{gray}\scriptsize$_{\pm35}$} & \textbf{+.0414} & \textbf{6.7\%} & \textbf{.7488}{\color{gray}\scriptsize$_{\pm18}$} & \textbf{+.0139} & \textbf{1.9\%} \\
\cmidrule{2-12}
& \multirow{4}{*}{HyFormer}
& Pos Seq & .8329{\color{gray}\scriptsize$_{\pm35}$} & --- & --- & .6201{\color{gray}\scriptsize$_{\pm28}$} & --- & --- & .7397{\color{gray}\scriptsize$_{\pm19}$} & --- & --- \\
& & Pos Seq + TAPF & .8384{\color{gray}\scriptsize$_{\pm27}$} & +.0055 & 0.7\% & .6208{\color{gray}\scriptsize$_{\pm21}$} & +.0007 & 0.1\% & .7413{\color{gray}\scriptsize$_{\pm24}$} & +.0016 & 0.2\% \\
& & Mix Seq + PB & .8745{\color{gray}\scriptsize$_{\pm21}$} & +.0416 & 5.0\% & .6720{\color{gray}\scriptsize$_{\pm26}$} & +.0519 & 8.4\% & .7532{\color{gray}\scriptsize$_{\pm17}$} & +.0135 & 1.8\% \\
& & Mix Seq + TAPF & \textbf{.8774}{\color{gray}\scriptsize$_{\pm26}$} & \textbf{+.0445} & \textbf{5.3\%} & \textbf{.6762}{\color{gray}\scriptsize$_{\pm22}$} & \textbf{+.0561} & \textbf{9.0\%} & \textbf{.7573}{\color{gray}\scriptsize$_{\pm14}$} & \textbf{+.0176} & \textbf{2.4\%} \\
\cmidrule{2-12}
& \multirow{4}{*}{MixFormer}
& Pos Seq & .8349{\color{gray}\scriptsize$_{\pm18}$} & --- & --- & .6162{\color{gray}\scriptsize$_{\pm28}$} & --- & --- & .7364{\color{gray}\scriptsize$_{\pm28}$} & --- & --- \\
& & Pos Seq + TAPF & .8377{\color{gray}\scriptsize$_{\pm22}$} & +.0028 & 0.3\% & .6197{\color{gray}\scriptsize$_{\pm24}$} & +.0035 & 0.6\% & .7401{\color{gray}\scriptsize$_{\pm25}$} & +.0037 & 0.5\% \\
& & Mix Seq + PB & .8724{\color{gray}\scriptsize$_{\pm26}$} & +.0375 & 4.5\% & .6609{\color{gray}\scriptsize$_{\pm34}$} & +.0447 & 7.3\% & .7514{\color{gray}\scriptsize$_{\pm25}$} & +.0150 & 2.0\% \\
& & Mix Seq + TAPF & \textbf{.8809}{\color{gray}\scriptsize$_{\pm25}$} & \textbf{+.0460} & \textbf{5.5\%} & \textbf{.6667}{\color{gray}\scriptsize$_{\pm27}$} & \textbf{+.0505} & \textbf{8.2\%} & \textbf{.7545}{\color{gray}\scriptsize$_{\pm24}$} & \textbf{+.0181} & \textbf{2.5\%} \\
\bottomrule
\end{tabular}
\end{table*}

\textbf{Observation 1: Universal improvement across paradigms.} Mixed-polarity sequences yield consistent improvements across \emph{both} the traditional two-stage (DIN, Transformer) and unified-block paradigms (OneTrans, HyFormer, MixFormer). The relative improvement ranges from +1.9\% to +9.6\%, with improvements observed in all 15 model--dataset combinations. This suggests that the value of negative behavioral context is largely independent of specific architectural design, representing a \emph{data-level} contribution that diverse models can exploit.

\textbf{Observation 2: TAPF consistently outperforms PB.} Across all models, Mix Seq + TAPF provides additional gains over the simpler Mix Seq + PB. On KuaiRec, MixFormer achieves 0.8809 vs.\ 0.8724 (+0.0085); HyFormer reaches 0.8774 vs.\ 0.8745 (+0.0029); DIN achieves 0.8711 vs.\ 0.8650 (+0.0061). The target-conditioned gating in TAPF provides finer-grained polarity differentiation than the uniform bias of PB. Notably, even on positive-only sequences (Pos Seq + TAPF), TAPF yields consistent improvements over the vanilla Pos Seq baseline across 14 out of 15 model--dataset combinations (+0.3\% to +0.7\% on KuaiRec), demonstrating that the target-conditioned gating mechanism is effective in its own right. However, these gains are substantially smaller than those achieved by introducing negative behaviors (Mix Seq), confirming that the primary source of improvement is the mixed-polarity data paradigm itself.

\begin{table}[t]
\centering
\caption{TAPF component ablation on KuaiRec (OneTrans, 3-seed mean). $\Delta$: AUC change relative to Full TAPF.}
\label{tab:tapf_ablation}
\small
\begin{tabular}{@{}l|ccc@{}}
\toprule
\textbf{Configuration} & \textbf{AUC} & $\Delta$ & $\Delta$\% \\
\midrule
Full TAPF & \textbf{.8845} & --- & --- \\
\midrule
$-$ Polarity embedding & .8842 & $-$.0003 & $-$0.03\% \\
$-$ Content-adaptive gating & .8838 & $-$.0007 & $-$0.08\% \\
$-$ Polarity sign ($m{=}\pm1$) & .8833 & $-$.0012 & $-$0.14\% \\
$-$ Residual connection & .8816 & $-$.0029 & $-$0.33\% \\
$-$ Target interaction & .8807 & $-$.0038 & $-$0.43\% \\
$-$ Target interaction \& gating & .8760 & $-$.0085 & $-$0.96\% \\
\midrule
PB only (no TAPF) & .8777 & $-$.0068 & $-$0.77\% \\
\bottomrule
\end{tabular}
\end{table}

To understand which components of TAPF contribute most, we conduct an ablation study on KuaiRec (OneTrans, 3-seed mean), shown in Table~\ref{tab:tapf_ablation}.The ablation reveals that \emph{target-aware interaction} is the most critical component ($-$0.0038 when removed), followed by the residual connection ($-$0.0029). Removing both target interaction and gating reduces TAPF to the level of PB only (0.8775 vs.\ 0.8777), confirming that target-conditioned feature transformation is the essential mechanism. The polarity sign, gating, and polarity embedding contribute progressively smaller increments, indicating that TAPF's effectiveness is primarily driven by its ability to compute target-relevant interaction features rather than by the gating or polarity encoding alone.

\subsection{RQ2: Sequence Length Scaling}

A central question for production systems is whether longer sequences justify their computational cost. We investigate how mixed-polarity sequences scale with increasing sequence length compared to positive-only sequences, using all five architectures on KuaiRec with lengths from 10 to 500. Table~\ref{tab:scaling} presents the full scaling results. For each model, we report AUC at each sequence length under both settings, the scaling slope ($\Delta_{10 \to 500}$), and the ratio of mixed-over-positive scaling efficiency.

\begin{table*}[t]
\centering
\caption{Sequence length scaling on KuaiRec. $\Delta_{10{\to}100}$: AUC gain from $L{=}10$ to $L{=}100$. $\Delta_{10{\to}500}$: AUC gain from $L{=}10$ to $L{=}500$. Ratio: slope improvement of Mix Seq + TAPF over Pos Seq.}
\label{tab:scaling}
\small
\begin{tabular}{@{}lll|ccccccc|c|c|c|c@{}}
\toprule
& & & \multicolumn{7}{c|}{\textbf{AUC at Sequence Length}} & & & & \\
\textbf{Type} & \textbf{Model} & \textbf{Setting} & $L{=}10$ & $L{=}20$ & $L{=}50$ & $L{=}100$ & $L{=}200$ & $L{=}300$ & $L{=}500$ & $\Delta_{10{\to}100}$ & Ratio & $\Delta_{10{\to}500}$ & Ratio \\
\midrule
\multirow{4}{*}{Two-Stage}
& \multirow{2}{*}{DIN}
& Pos Seq & .8263 & .8306 & .8351 & .8362 & .8374 & .8366 & .8365 & +.0099 & \multirow{2}{*}{\textbf{1.97$\times$}} & +.0102 & \multirow{2}{*}{\textbf{2.15$\times$}} \\
& & Mix Seq + TAPF & .8516 & .8614 & .8688 & .8711 & .8733 & .8743 & .8735 & +.0195 & & +.0219 & \\
\cmidrule{2-14}
& \multirow{2}{*}{Transformer}
& Pos Seq & .8302 & .8364 & .8440 & .8454 & .8472 & .8466 & .8470 & +.0152 & \multirow{2}{*}{\textbf{1.90$\times$}} & +.0168 & \multirow{2}{*}{\textbf{1.90$\times$}} \\
& & Mix Seq + TAPF & .8519 & .8644 & .8752 & .8808 & .8840 & .8841 & .8839 & +.0289 & & +.0320 & \\
\midrule
\multirow{6}{*}[-0.5em]{Unified Block}
& \multirow{2}{*}{OneTrans}
& Pos Seq & .8323 & .8403 & .8457 & .8462 & .8466 & .8469 & .8472 & +.0139 & \multirow{2}{*}{\textbf{1.86$\times$}} & +.0149 & \multirow{2}{*}{\textbf{1.93$\times$}} \\
& & Mix Seq + TAPF & .8587 & .8695 & .8776 & .8845 & .8867 & .8872 & .8875 & +.0258 & & +.0288 & \\
\cmidrule{2-14}
& \multirow{2}{*}{HyFormer}
& Pos Seq & .8196 & .8255 & .8322 & .8329 & .8338 & .8340 & .8345 & +.0133 & \multirow{2}{*}{\textbf{2.08$\times$}} & +.0149 & \multirow{2}{*}{\textbf{2.13$\times$}} \\
& & Mix Seq + TAPF & .8498 & .8620 & .8711 & .8774 & .8805 & .8810 & .8815 & +.0276 & & +.0317 & \\
\cmidrule{2-14}
& \multirow{2}{*}{MixFormer}
& Pos Seq & .8201 & .8284 & .8335 & .8349 & .8357 & .8362 & .8366 & +.0148 & \multirow{2}{*}{\textbf{2.34$\times$}} & +.0165 & \multirow{2}{*}{\textbf{2.22$\times$}} \\
& & Mix Seq + TAPF & .8462 & .8612 & .8744 & .8809 & .8822 & .8825 & .8829 & +.0347 & & +.0367 & \\
\bottomrule
\end{tabular}
\end{table*}

The scaling experiment, visualized in Figure~\ref{fig:teaser} (page 1), reveals several key findings:

\textbf{Observation 1: Mixed sequences achieve $\sim$2$\times$ steeper scaling slopes.} Across all five architectures, the scaling slope of mixed-polarity sequences from $L{=}10$ to $L{=}500$ is approximately double that of positive-only sequences. The ratio ranges from 1.90$\times$ (Transformer) to 2.22$\times$ (MixFormer), and this advantage holds equally for the traditional two-stage models (DIN: 2.15$\times$, Transformer: 1.90$\times$) and unified-block models, suggesting that the improved scaling is primarily a property of the \emph{data composition} rather than a specific architectural mechanism.

\begin{figure*}[t]
\centering
\begin{subfigure}[b]{0.28\textwidth}
\includegraphics[width=\textwidth]{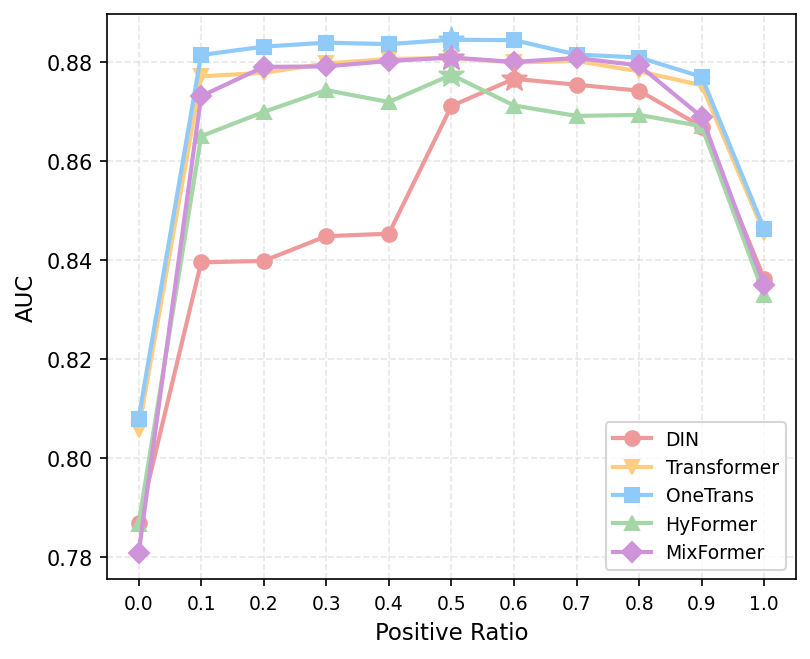}
\caption{KuaiRec}
\end{subfigure}\hfill
\begin{subfigure}[b]{0.28\textwidth}
\includegraphics[width=\textwidth]{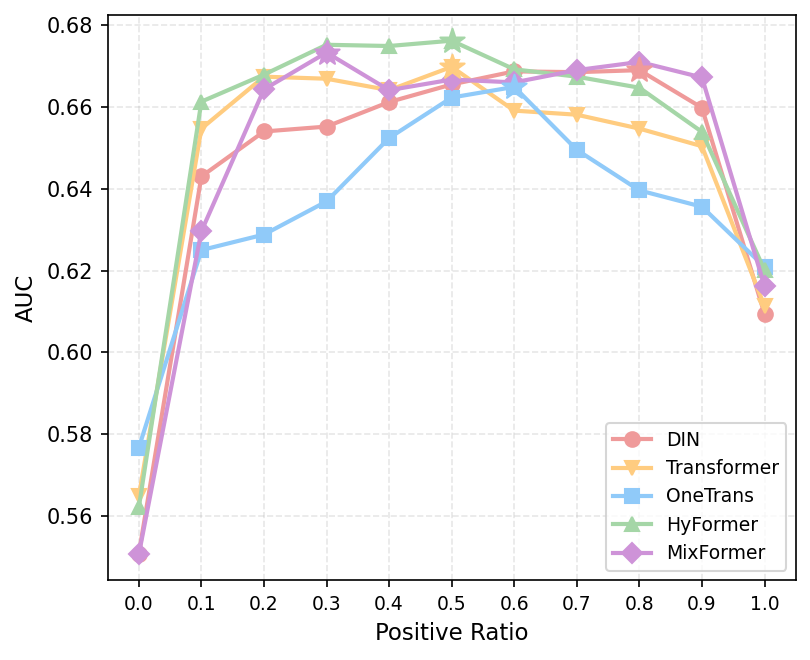}
\caption{KuaiRand}
\end{subfigure}\hfill
\begin{subfigure}[b]{0.28\textwidth}
\includegraphics[width=\textwidth]{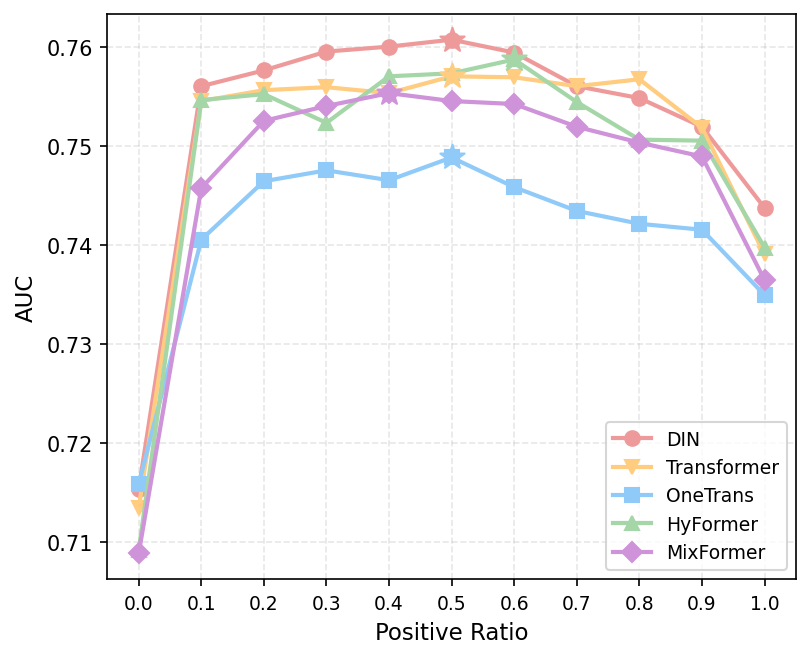}
\caption{TAAC}
\end{subfigure}
\caption{AUC vs.\ positive ratio $r$ across datasets. All curves exhibit an inverted-U shape with optimal $r^* \in [0.2, 0.7]$ and a broad performance plateau.}
\label{fig:ratio}
\end{figure*}

\textbf{Observation 2: Positive-only sequences plateau early due to inherent length constraints.} The dashed curves in Figure~\ref{fig:teaser} flatten noticeably after $L{\approx}50$--100. For example, OneTrans (Pos Seq) gains only +0.0010 from $L{=}100$ to $L{=}500$ (0.8462$\to$0.8472). A key contributing factor is that users' positive behavior histories are inherently bounded by their actual engagement rate; once the sequence length exceeds the available positive interactions, further extension yields diminishing information. In contrast, mixed-polarity sequences draw from a substantially larger pool (positive + negative histories combined), providing both richer information content and greater room for continued scaling. OneTrans (Mix Seq) gains +0.0030 over the same range (0.8845$\to$0.8875), maintaining meaningful improvement even at long lengths.

\textbf{Observation 3: Signal quality, not data sufficiency, drives the improvement.} A potential confound in the above analysis is that positive-only sequences may suffer from insufficient data (padding) for users with short histories, while mixed sequences benefit from filling slots with available negatives. To isolate signal quality from data sufficiency, we evaluate on the \emph{full-sequence subset} of KuaiRec: test samples whose positive history length $\geq 100$, ensuring that both Pos Seq and Mix Seq are fully filled without padding (28.6\% of test samples). Table~\ref{tab:fullseq} shows that on this subset, the scaling ratio of Mix Seq + TAPF over Pos Seq remains 1.82--2.10$\times$, only slightly below the 1.90--2.22$\times$ observed on the full test set. This confirms that the advantage of mixed-polarity sequences is predominantly driven by the superior signal quality of negative behavioral information, rather than merely filling otherwise empty sequence slots.

\begin{table}[t]
\centering
\caption{Scaling on KuaiRec full-sequence subset (positive history $\geq$100). Both settings are fully filled without padding. Ratio: slope improvement of Mix Seq + TAPF over Pos Seq.}
\label{tab:fullseq}
\small
\begin{tabular}{@{}ll|ccc|c|c@{}}
\toprule
\textbf{Model} & \textbf{Setting} & $L{=}10$ & $L{=}50$ & $L{=}100$ & $\Delta_{10{\to}100}$ & Ratio \\
\midrule
\multirow{2}{*}{DIN}
& Pos Seq & .7833 & .7925 & .7938 & +.0105 & \multirow{2}{*}{\textbf{2.01$\times$}} \\
& Mix + TAPF & .8194 & .8374 & .8405 & +.0211 & \\
\midrule
\multirow{2}{*}{Transformer}
& Pos Seq & .7830 & .7956 & .7969 & +.0139 & \multirow{2}{*}{\textbf{1.83$\times$}} \\
& Mix + TAPF & .8243 & .8422 & .8498 & +.0255 & \\
\midrule
\multirow{2}{*}{OneTrans}
& Pos Seq & .7878 & .7994 & .8019 & +.0141 & \multirow{2}{*}{\textbf{1.82$\times$}} \\
& Mix + TAPF & .8268 & .8455 & .8525 & +.0257 & \\
\midrule
\multirow{2}{*}{HyFormer}
& Pos Seq & .7867 & .7964 & .7981 & +.0114 & \multirow{2}{*}{\textbf{2.04$\times$}} \\
& Mix + TAPF & .8232 & .8400 & .8464 & +.0232 & \\
\midrule
\multirow{2}{*}{MixFormer}
& Pos Seq & .7796 & .7913 & .7945 & +.0149 & \multirow{2}{*}{\textbf{2.10$\times$}} \\
& Mix + TAPF & .8185 & .8408 & .8498 & +.0313 & \\
\bottomrule
\end{tabular}
\end{table}

\begin{figure*}[t]
\centering
\begin{subfigure}[b]{0.305\textwidth}
\includegraphics[width=\textwidth]{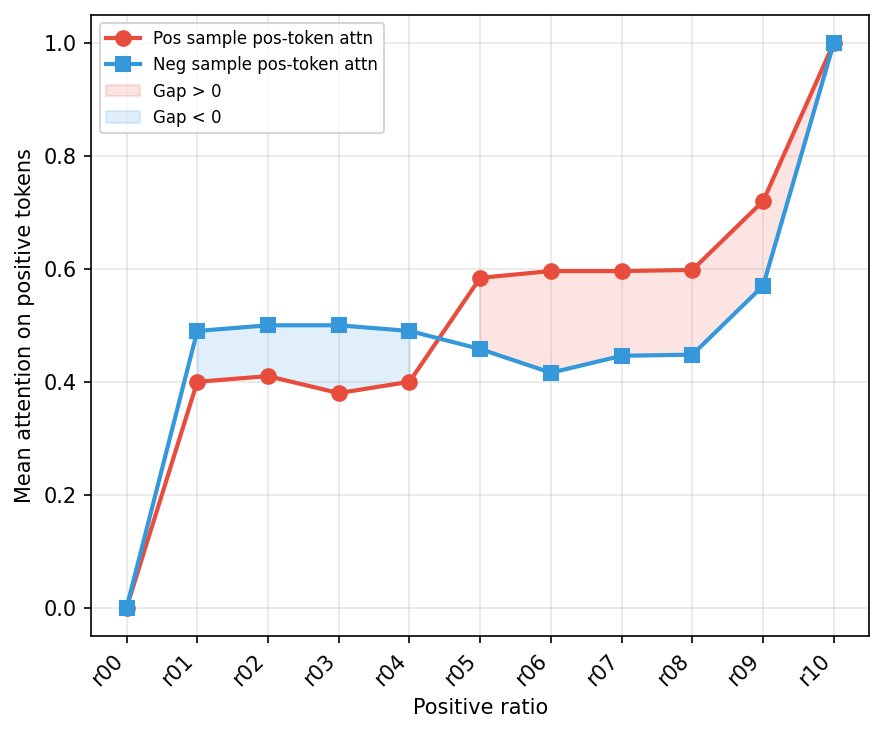}
\caption{Attention gap}
\end{subfigure}\hfill
\begin{subfigure}[b]{0.375\textwidth}
\includegraphics[width=\textwidth]{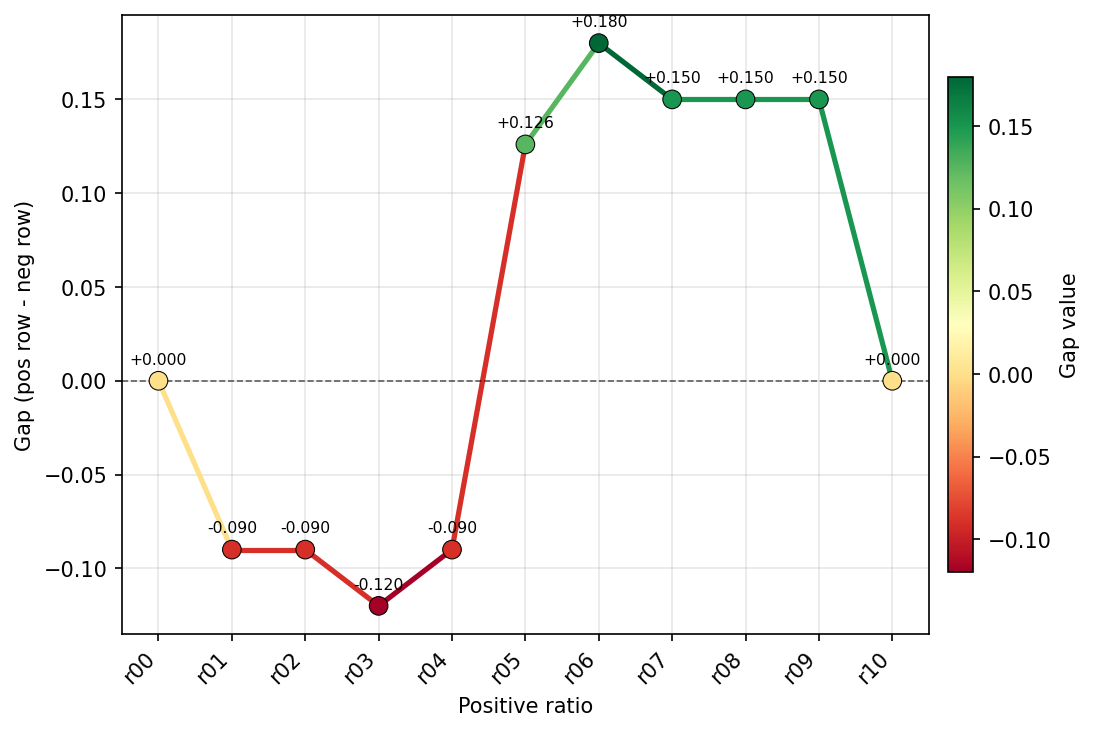}
\caption{Gap trend}
\end{subfigure}\hfill
\begin{subfigure}[b]{0.305\textwidth}
\includegraphics[width=\textwidth]{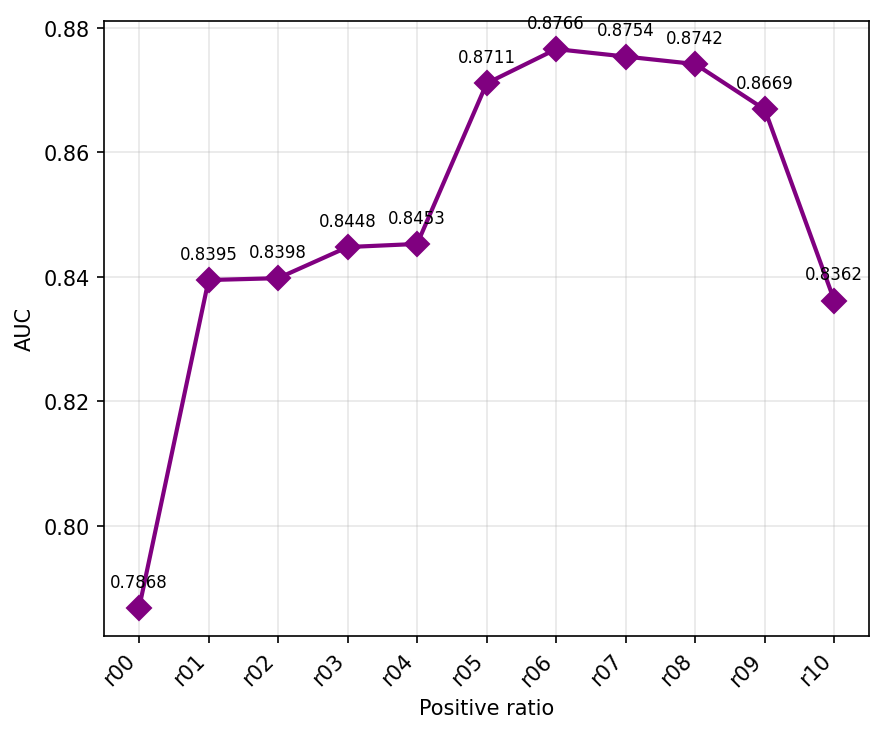}
\caption{AUC trend}
\end{subfigure}
\caption{KuaiRec (DIN): attention polarity gap vs.\ positive ratio. The gap transitions from negative to positive at $r{\approx}0.5$, precisely where AUC jumps.}
\label{fig:attn_kuairec}
\end{figure*}

\subsection{RQ3: Positive-Negative Composition Ratio}

We systematically vary the positive ratio $r \in \{0.0, 0.1, \ldots, 1.0\}$, controlling what fraction of the fixed-length sequence budget is allocated to positive tokens. Figure~\ref{fig:ratio} reveals a remarkably consistent pattern across all configurations.

\begin{table}[t]
\centering
\caption{Attention polarity gap at $r{=}0.5$.}
\label{tab:attn_gap}
\small
\begin{tabular}{@{}l|cc|c@{}}
\toprule
\textbf{Dataset (Model)} & Pos$\to$PosToken & Neg$\to$PosToken & Gap \\
\midrule
KuaiRec (OneTrans) & 0.659 & 0.611 & +0.048 \\
KuaiRec (DIN) & 0.584 & 0.458 & +0.125 \\
TAAC (OneTrans) & 0.236 & 0.109 & +0.128 \\
\bottomrule
\end{tabular}
\end{table}

\textbf{Inverted-U shape with broad plateau.} Every model--dataset combination produces a curve that rises from $r{=}0.0$, reaches a plateau in the middle range, and declines toward $r{=}1.0$. The consistency of this shape across three diverse datasets and architecturally distinct models suggests that mixed-polarity modeling captures a general characteristic of user behavior rather than an artifact of any specific data distribution. The vast majority of optimal ratios fall within $r^* \in [0.2, 0.7]$, and performance within this range is remarkably stable: on KuaiRec, OneTrans varies by less than 0.005 AUC across $r \in [0.2, 0.7]$. This robustness is practically important, as it means a default of $r{=}0.5$ captures the majority of available improvement in all tested scenarios without requiring expensive hyperparameter search.

\begin{figure}[t]
\centering
\begin{subfigure}[b]{0.75\columnwidth}
\includegraphics[width=\textwidth]{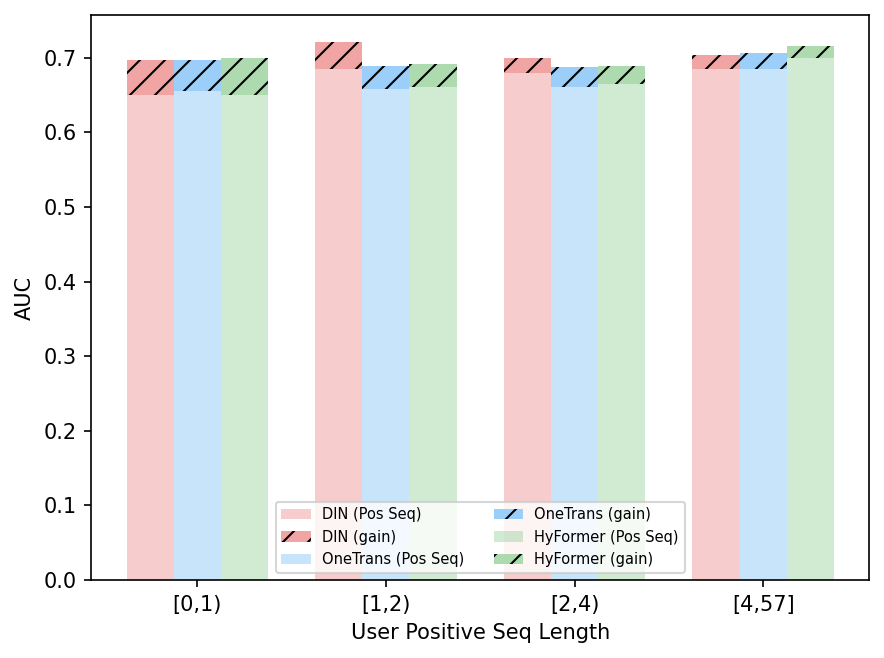}
\caption{TAAC}
\end{subfigure}\\[0.3em]
\begin{subfigure}[b]{0.75\columnwidth}
\includegraphics[width=\textwidth]{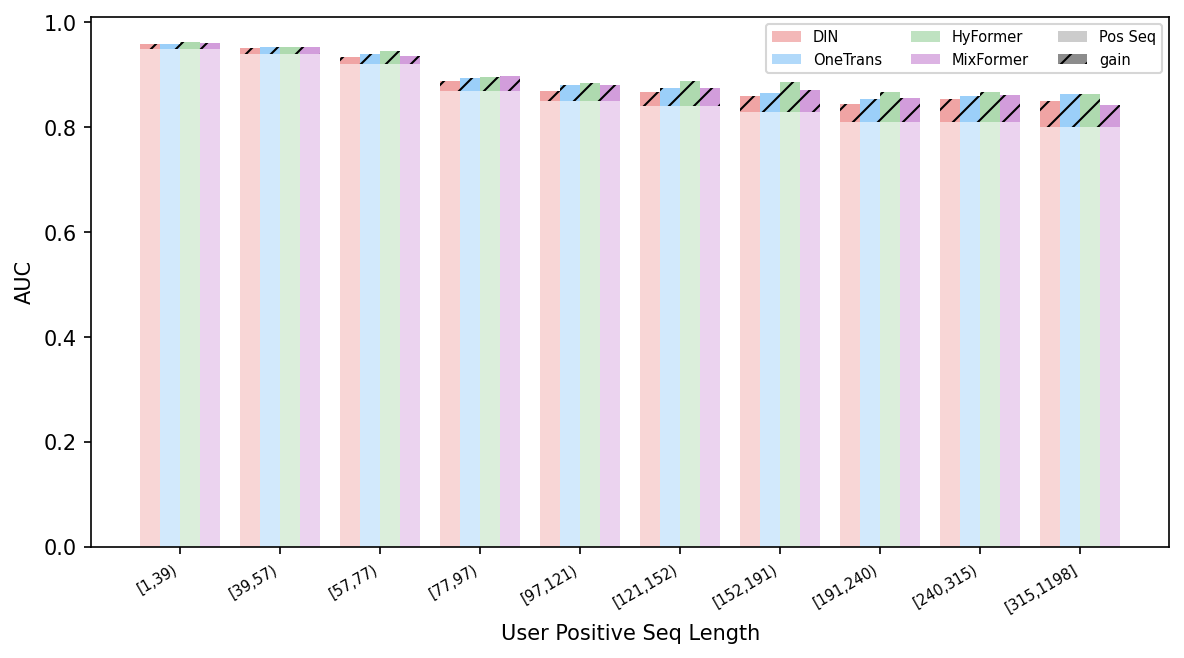}
\caption{KuaiRec}
\end{subfigure}
\caption{User activity stratification. (a)~TAAC: low-activity users (0--1 positive behaviors) gain 3--4$\times$ more. (b)~KuaiRec: high-activity users gain more due to the complementary information mechanism.}
\label{fig:user_activity}
\end{figure}

\subsection{RQ4: Why Do Mixed-Polarity Sequences Work?}

We investigate the mechanisms through six complementary analyses, each addressing a specific question about how and why negative behaviors contribute to prediction quality.

\subsubsection{How does the model's attention change?}

We observe that the DIN model exhibits a sharp AUC jump at positive ratio $r{\approx}0.5$ on KuaiRec (Figure~\ref{fig:ratio}a), which draws our attention to the internal mechanism at this transition point. We define the \textbf{attention polarity gap} as the difference in attention allocated to positive tokens between positive and negative test samples ($\text{gap} = \mathbb{E}[\bar{a}^+ | y{=}1] - \mathbb{E}[\bar{a}^+ | y{=}0]$). A non-zero gap (either positive or negative) indicates that the model exhibits differentiated attention allocation patterns across positive and negative test samples, reflecting its ability to discriminate between them.

Table~\ref{tab:attn_gap} and Figure~\ref{fig:attn_kuairec} reveal that the model spontaneously learns polarity-discriminative attention: positive test samples allocate more weight to positive history tokens, while negative samples shift attention toward negative tokens. Crucially, the gap undergoes a sharp phase transition from negative to positive at $r{\approx}0.5$ (Figure~\ref{fig:attn_kuairec}b), which precisely coincides with the AUC jump at the same threshold (Figure~\ref{fig:attn_kuairec}c). This strong correlation suggests that the AUC improvement is closely associated with the model's ability to differentially attend to polarity-relevant tokens.

\subsubsection{Are all users equally affected?}

We stratify users by positive sequence length to test whether gains are uniform:

\textbf{TAAC} (Figure~\ref{fig:user_activity}a): Users with 0--1 positive behaviors gain +0.041 to +0.050 AUC, versus +0.013 to +0.021 for users with 4--57 behaviors. For data-sparse users, negative tokens supply \emph{foundational information}: the model transitions from ``knowing nothing'' to ``knowing what the user dislikes,'' providing a strong signal in an otherwise void representation.

\textbf{KuaiRec} (Figure~\ref{fig:user_activity}b): Unlike TAAC where user positive histories are short (max 57, Table~\ref{tab:datasets}), KuaiRec features much longer positive histories (max 2,244, Table~\ref{tab:datasets}), meaning most users have positive behaviors far exceeding the length budget $L{=}100$. Here the pattern reverses: high-activity users (315--1200 positive behaviors) gain +0.042 to +0.063, while low-activity users (1--39) gain only +0.004 to +0.012. This indicates that for data-rich users, replacing outdated positive behaviors with recent negative behaviors within the same length budget yields better representations, as recent dispreference signals carry more informative value than stale positive interactions.

These two datasets reveal that negative behaviors serve dual roles: \emph{foundational} for data-sparse users and \emph{complementary} for data-rich users whose positive sequences are truncated.

\subsubsection{Do cold-start items benefit?}

\begin{figure}[t]
\centering
\begin{subfigure}[b]{0.48\columnwidth}
\includegraphics[width=\textwidth]{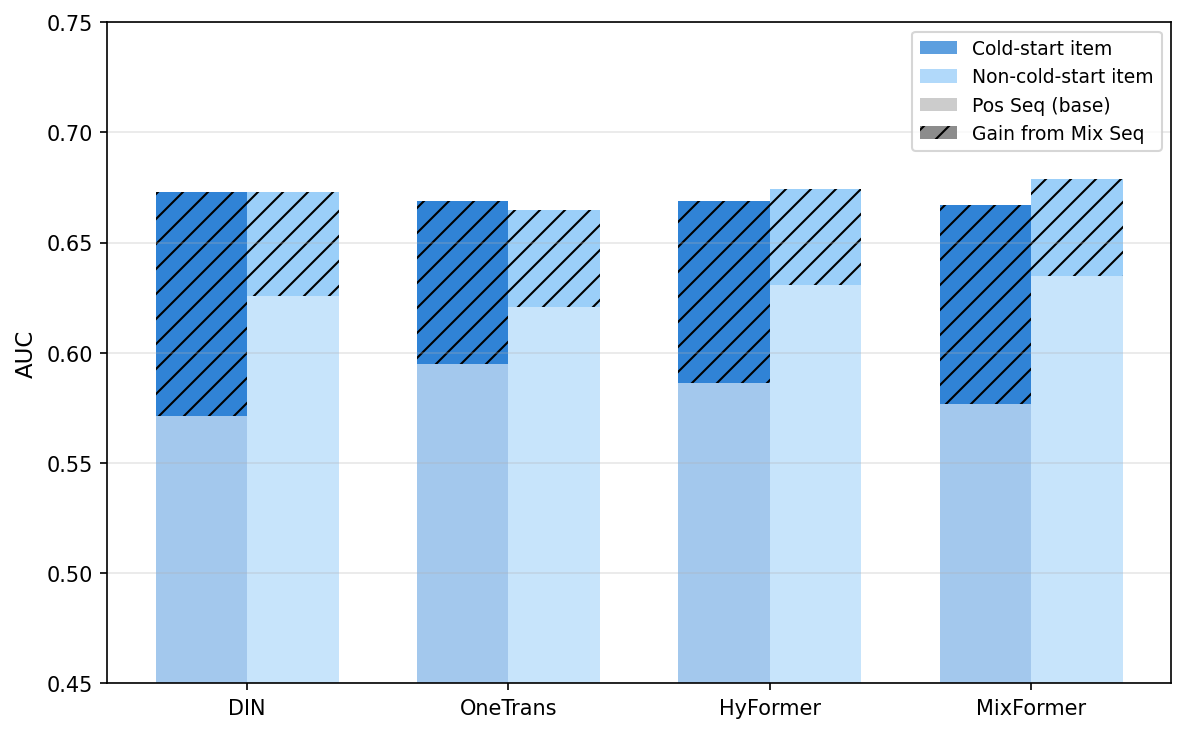}
\caption{KuaiRand}
\end{subfigure}\hfill
\begin{subfigure}[b]{0.48\columnwidth}
\includegraphics[width=\textwidth]{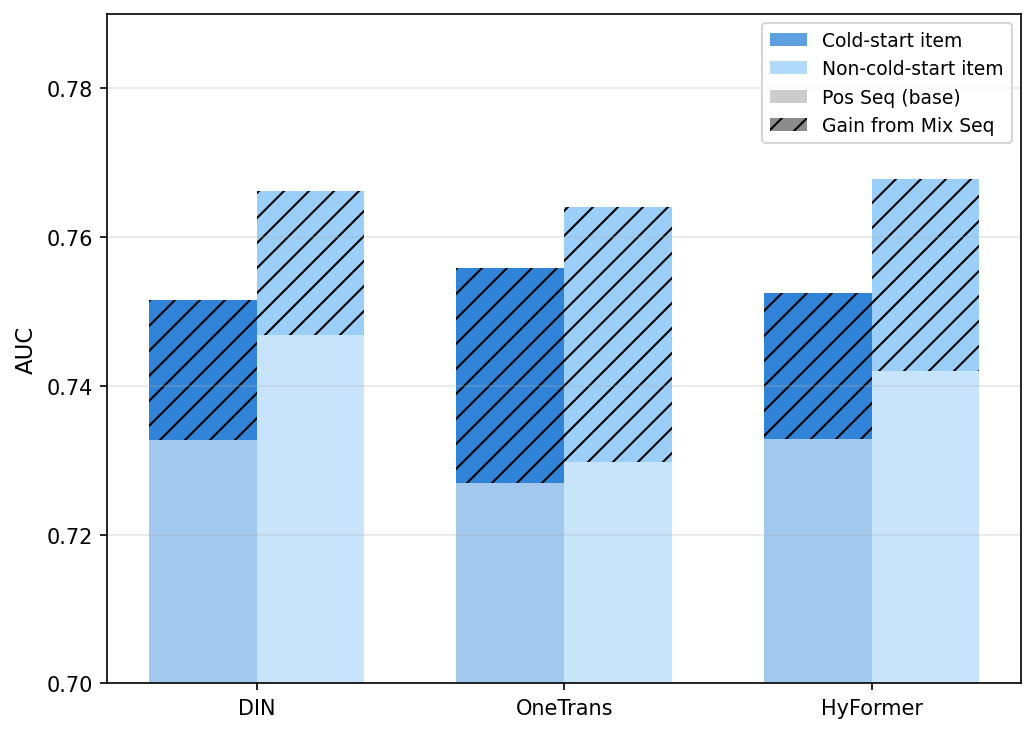}
\caption{TAAC}
\end{subfigure}
\caption{Cold-start item analysis. For each model, left bar: cold-start items; right bar: non-cold items. Lighter portion: Pos Seq baseline; hatched: gain from Mix Seq. Cold items gain 1.5--2$\times$ more on KuaiRand; gains are comparable on TAAC.}
\label{fig:cold_start}
\end{figure}

We evaluate items with zero training-set exposure (cold-start) on KuaiRand (30.1\% cold items) and TAAC (25.2\% cold items). Figure~\ref{fig:cold_start} shows:

On KuaiRand, cold-start items gain +0.074 to +0.102 AUC from mixed sequences, compared to +0.043 to +0.047 for non-cold items (1.5--2$\times$ larger). When target item embeddings are untrained, the model cannot rely on item-level representations; negative tokens in the user's sequence provide \emph{indirect exclusion signals} that constrain the prediction space without requiring item-specific knowledge.

On TAAC, the gains for cold-start items and non-cold items are largely comparable, suggesting that the cold-start advantage is more pronounced when the embedding space is sufficiently expressive.

\subsubsection{Does the ``negative'' label matter?}

\begin{figure}[t]
\centering
\begin{subfigure}[b]{0.32\columnwidth}
\includegraphics[width=\textwidth]{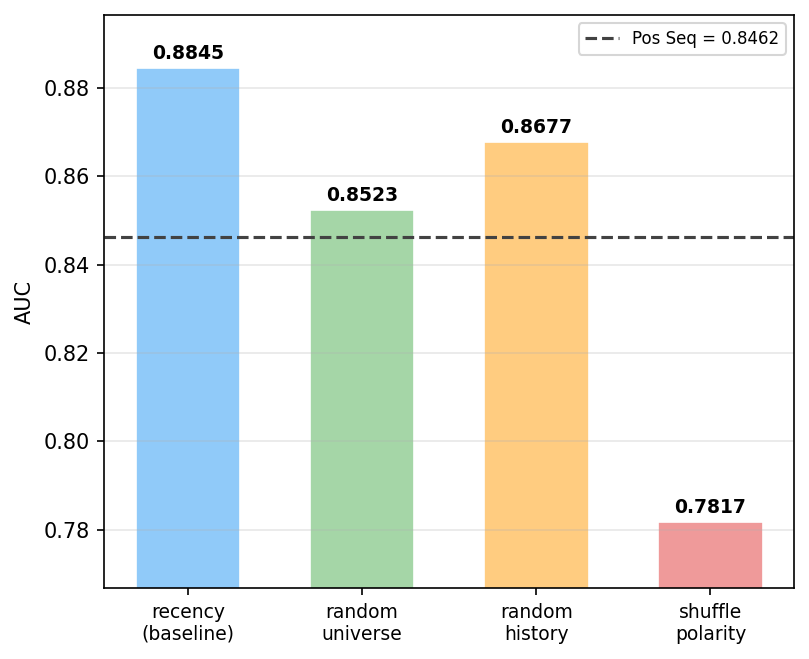}
\caption{KuaiRec}
\end{subfigure}\hfill
\begin{subfigure}[b]{0.32\columnwidth}
\includegraphics[width=\textwidth]{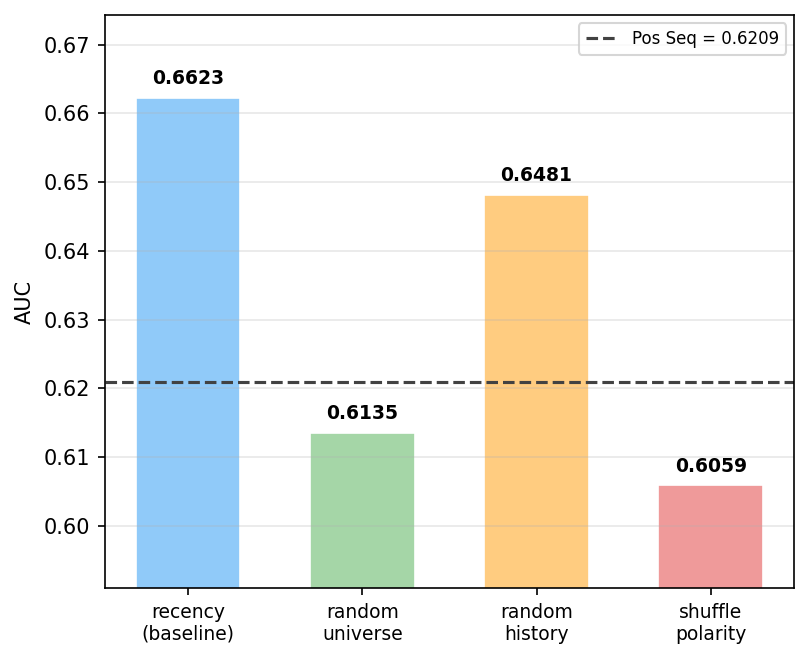}
\caption{KuaiRand}
\end{subfigure}\hfill
\begin{subfigure}[b]{0.32\columnwidth}
\includegraphics[width=\textwidth]{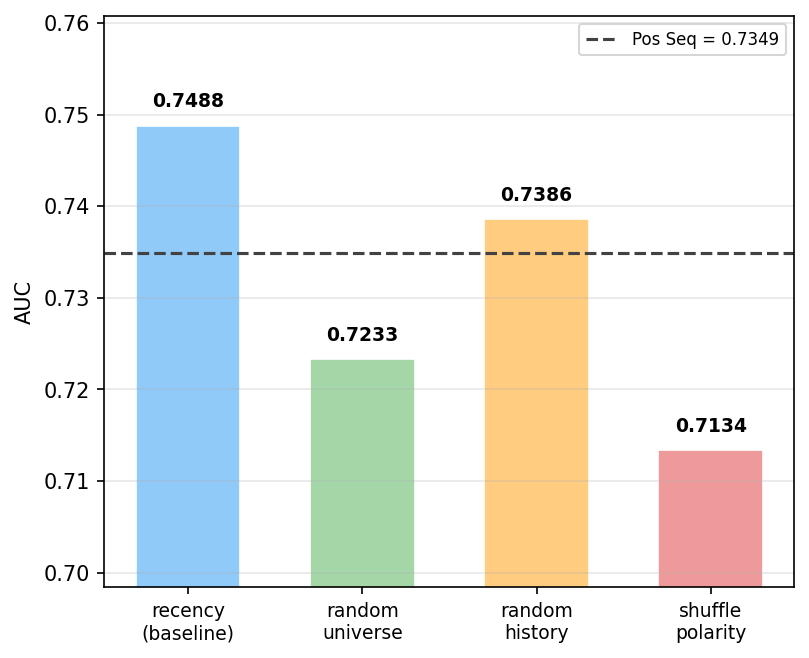}
\caption{TAAC}
\end{subfigure}
\caption{Negative behavior ablation. Real negatives (recency) consistently outperform random alternatives.}
\label{fig:neg_ablation}
\end{figure}

To disentangle the sources of improvement in mixed-polarity sequences, we design three ablation conditions that isolate distinct factors (Figure~\ref{fig:neg_ablation}):

\textbf{(a) Does the improvement require real negative behaviors?} We replace negative tokens with \textbf{random universe} (globally sampled random items unrelated to the user). On KuaiRec (10K items), random universe slightly outperforms Pos Seq, suggesting that sequence diversity provides marginal benefits when the item space is small. However, on KuaiRand (4.37M items) and TAAC (2.3M items), random universe \emph{underperforms} Pos Seq, indicating that randomly sampled items from a large vocabulary introduce predominantly irrelevant noise. Since real-world industrial systems typically operate over millions of items, this finding suggests that the diversity brought by arbitrary item injection is unlikely to be the source of improvement in practice; the gains from mixed-polarity sequences stem from genuine dispreference semantics carried by real negative behaviors.

\textbf{(b) Does recency matter?} We replace recent negatives with \textbf{random history} (randomly selected negative behaviors from the user's full history, ignoring recency). Recency consistently outperforms random history across all datasets, confirming that recent negative behaviors carry stronger relevance signals: items recently skipped are more likely to be semantically related to the user's current interests, providing more informative dispreference context than temporally distant negative interactions.

\textbf{(c) Do accurate polarity labels matter?} We construct the mixed sequence with correct items but \textbf{randomly assigned polarity labels} (random polarity). This condition produces the most severe degradation across all datasets, falling substantially below even the positive-only baseline. This demonstrates that accurate polarity labels play a decisive role: when positive and negative behaviors are assigned random polarities, the model receives contradictory semantic signals that actively corrupt the user representation, leading to performance far worse than simply ignoring negative behaviors altogether.

\textbf{Summary.} These ablations collectively indicate that accurate, genuine negative behaviors are the key driver of performance improvement. Additionally, recency contributes a meaningful portion of the gains: selecting the most recent negative behaviors yields further improvement over random historical negatives, suggesting that temporal proximity enhances the informativeness of dispreference signals.

\subsubsection{What does TAPF change mechanistically?}

\begin{figure}[t]
\centering
\includegraphics[width=0.55\columnwidth]{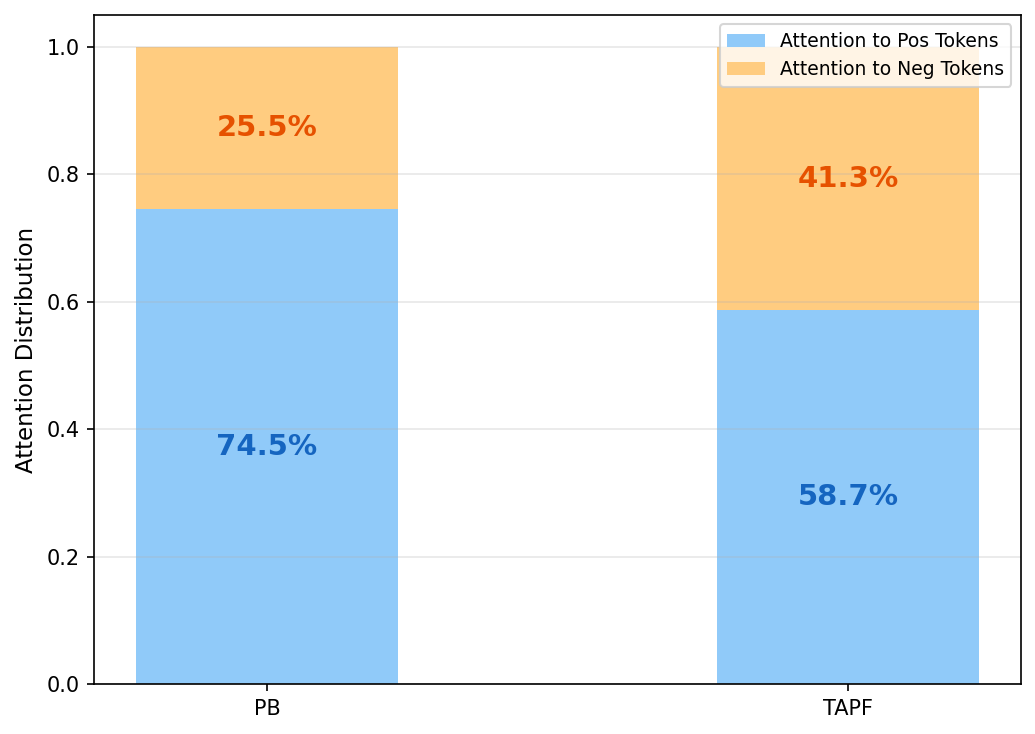}
\caption{TAPF vs.\ PBE attention distribution on KuaiRec (OneTrans). Each bar shows the split between attention to positive tokens (bottom, blue) and negative tokens (top, orange). TAPF rebalances from 74.5\%:25.5\% to 58.7\%:41.3\%.}
\label{fig:tapf}
\end{figure}

Finally, we analyze how TAPF changes the model's internal behavior. Using the same attention computation as in the attention polarity gap analysis (\S4.5.1), we measure the proportion of attention allocated to positive versus negative tokens. On KuaiRec (OneTrans), PBE allocates 74.5\% of attention to positive tokens, effectively underutilizing negative behaviors. TAPF rebalances this to 58.7\%:41.3\% (Figure~\ref{fig:tapf}), a 62\% increase in negative token utilization. This indicates that TAPF achieves more effective polarity-aware semantic calibration, enabling the model to more fully leverage information carried by negative behaviors. The increased attention to negative tokens, combined with an AUC improvement of +0.0068 (from 0.8777 to 0.8845), suggests that TAPF's target-conditioned gating better utilizes the informational value of negative behaviors compared to PBE's uniform bias.

\subsubsection{Why a mixed sequence, not separate streams?}

A natural alternative to chronologically interleaved mixed sequences is to process positive and negative behaviors separately. We compare against two such alternatives on KuaiRec (OneTrans, 5-seed mean):

\textbf{Mixed $>$ Dual-Stream} ($-$0.0265): Dual-stream processes positive and negative histories with separate encoders, fusing their outputs only at the prediction stage. Its underperformance demonstrates that \emph{cross-polarity interaction} during encoding is valuable: the model benefits from jointly attending to positive and negative behaviors within the same attention computation, capturing the interplay between what a user likes and dislikes that dual-stream architectures cannot model.

\textbf{Mixed $>$ Polarity-Grouped Concat} ($-$0.0216): This approach concatenates all positive tokens first, then all negative tokens, within a single encoder. Its degradation shows that the chronological interleaving of positive and negative behaviors carries a unique \emph{alternation signal}: the pattern of ``skip-then-click'' or ``click-then-skip'' encodes behavioral context that is lost when tokens are grouped by polarity rather than ordered by time.

\section{Conclusion}

We have demonstrated that implicit negative behaviors, when incorporated into user behavior sequences via chronological interleaving and polarity-aware encoding, provide consistent and significant improvements to CTR prediction across diverse architectures. The Target-Aware Polarity Fusion mechanism addresses semantic indistinguishability by rebalancing attention toward negative tokens. Mechanistic analyses reveal that the improvements arise from polarity-discriminative attention, benefit both data-sparse and data-rich users through complementary mechanisms, and extend to cold-start items. Ablation studies confirm that genuine negative behaviors carry information beyond mere sequence diversity, and that chronological interleaving outperforms alternative sequence construction strategies. The framework is model-agnostic and lightweight, suggesting that mixed-polarity sequential modeling is a promising and broadly applicable direction for recommendation systems.

\begin{table}[t]
\centering
\caption{Sequence construction ablation on KuaiRec (OneTrans, 5-seed mean). Both alternatives underperform the chronologically interleaved mixed sequence.}
\label{tab:ablation}
\small
\begin{tabular}{@{}l|cc@{}}
\toprule
\textbf{Configuration} & \textbf{AUC} & $\Delta$ vs.\ Mixed \\
\midrule
Mixed Seq (chronological) & \textbf{.8802} & --- \\
Polarity-Grouped Concat & .8586 & $-$.0216 \\
Dual-Stream Encoder & .8537 & $-$.0265 \\
\bottomrule
\end{tabular}
\end{table}

\bibliographystyle{ACM-Reference-Format}
\bibliography{references}

@inproceedings{zhou2018din,
  author    = {Zhou, Guorui and Zhu, Xiaoqiang and Song, Chenru and Fan, Ying and Zhu, Han and Ma, Xiao and Yan, Yanghui and Jin, Junqi and Li, Han and Gai, Kunlin},
  title     = {Deep Interest Network for Click-Through Rate Prediction},
  booktitle = {Proceedings of the 24th ACM SIGKDD International Conference on Knowledge Discovery and Data Mining},
  pages     = {1059--1068},
  year      = {2018}
}

@inproceedings{zhou2019dien,
  author    = {Zhou, Guorui and Mou, Na and Fan, Ying and Pi, Qi and Bian, Weijie and Zhou, Chang and Zhu, Xiaoqiang and Gai, Kunlin},
  title     = {Deep Interest Evolution Network for Click-Through Rate Prediction},
  booktitle = {Proceedings of the 33rd AAAI Conference on Artificial Intelligence},
  pages     = {5941--5948},
  year      = {2019}
}

@inproceedings{pi2020search,
  author    = {Pi, Qi and Zhou, Guorui and Zhang, Yujing and Wang, Zhe and Ren, Lejian and Fan, Ying and Zhu, Xiaoqiang and Gai, Kunlin},
  title     = {Search-based User Interest Modeling with Lifelong Sequential Behavior Data for Click-Through Rate Prediction},
  booktitle = {Proceedings of the 29th ACM International Conference on Information and Knowledge Management},
  pages     = {2685--2692},
  year      = {2020}
}

@inproceedings{pi2019practice,
  author    = {Pi, Qi and Bian, Weijie and Zhou, Guorui and Zhu, Xiaoqiang and Gai, Kunlin},
  title     = {Practice on Long Sequential User Behavior Modeling for Click-Through Rate Prediction},
  booktitle = {Proceedings of the 25th ACM SIGKDD International Conference on Knowledge Discovery and Data Mining},
  pages     = {2671--2679},
  year      = {2019}
}

@inproceedings{chang2023twin,
  author    = {Chang, Jianxin and Zhang, Chenbin and Hui, Yiqun and Leng, Dewei and Liao, Mingming and He, Weijie and Gai, Kunlin and Ding, Ruobing and Jing, Biao},
  title     = {{TWIN}: TWo-stage Interest Network for Lifelong User Behavior Modeling in CTR Prediction at Kuaishou},
  booktitle = {Proceedings of the 29th ACM SIGKDD International Conference on Knowledge Discovery and Data Mining},
  pages     = {3785--3794},
  year      = {2023}
}

@inproceedings{chen2019behavior,
  author    = {Chen, Qiwei and Zhao, Huan and Li, Wei and Huang, Pipei and Ou, Wenwu},
  title     = {Behavior Sequence Transformer for E-commerce Recommendation in Alibaba},
  booktitle = {Proceedings of the 1st International Workshop on Deep Learning Practice for High-Dimensional Sparse Data (DLP-KDD)},
  year      = {2019}
}

@inproceedings{sun2019bert4rec,
  author    = {Sun, Fei and Liu, Jun and Wu, Jian and Pei, Changhua and Lin, Xiao and Ou, Wenwu and Jiang, Peng},
  title     = {{BERT4Rec}: Sequential Recommendation with Bidirectional Encoder Representations from Transformer},
  booktitle = {Proceedings of the 28th ACM International Conference on Information and Knowledge Management},
  pages     = {1441--1450},
  year      = {2019}
}

@inproceedings{guo2017deepfm,
  author    = {Guo, Huifeng and Tang, Ruiming and Ye, Yunming and Li, Zhenguo and He, Xiuqiang},
  title     = {{DeepFM}: A Factorization-Machine based Neural Network for CTR Prediction},
  booktitle = {Proceedings of the 26th International Joint Conference on Artificial Intelligence},
  pages     = {1725--1731},
  year      = {2017}
}

@inproceedings{cheng2016wide,
  author    = {Cheng, Heng-Tze and Koc, Levent and Harmsen, Jeremiah and Shaked, Tal and Chandra, Tushar and Aradhye, Hrishi and Anderson, Glen and Corrado, Greg and Chai, Wei and Ispir, Mustafa and others},
  title     = {Wide \& Deep Learning for Recommender Systems},
  booktitle = {Proceedings of the 1st Workshop on Deep Learning for Recommender Systems},
  pages     = {7--10},
  year      = {2016}
}

@inproceedings{wang2017deep,
  author    = {Wang, Ruoxi and Fu, Bin and Fu, Gang and Wang, Mingliang},
  title     = {Deep \& Cross Network for Ad Click Predictions},
  booktitle = {Proceedings of the ADKDD Workshop at KDD},
  year      = {2017}
}

@inproceedings{wang2021dcnv2,
  author    = {Wang, Ruoxi and Shivanna, Rakesh and Cheng, Derek and Jain, Sagar and Lin, Dong and Hong, Lichan and Chi, Ed},
  title     = {{DCN V2}: Improved Deep \& Cross Network and Practical Lessons for Web-scale Learning to Rank Systems},
  booktitle = {Proceedings of the Web Conference 2021},
  pages     = {1785--1797},
  year      = {2021}
}

@inproceedings{rendle2010fm,
  author    = {Rendle, Steffen},
  title     = {Factorization Machines},
  booktitle = {Proceedings of the 10th IEEE International Conference on Data Mining},
  pages     = {995--1000},
  year      = {2010}
}

@inproceedings{richardson2007predicting,
  author    = {Richardson, Matthew and Dominowska, Ewa and Ragno, Robert},
  title     = {Predicting Clicks: Estimating the Click-Through Rate for New Ads},
  booktitle = {Proceedings of the 16th International Conference on World Wide Web},
  pages     = {521--530},
  year      = {2007}
}

@inproceedings{rendle2009bpr,
  author    = {Rendle, Steffen and Freudenthaler, Christoph and Gantner, Zeno and Schmidt-Thieme, Lars},
  title     = {{BPR}: Bayesian Personalized Ranking from Implicit Feedback},
  booktitle = {Proceedings of the 25th Conference on Uncertainty in Artificial Intelligence},
  pages     = {452--461},
  year      = {2009}
}

@inproceedings{zhang2013optimizing,
  author    = {Zhang, Weinan and Chen, Tianqi and Wang, Jun and Yu, Yong},
  title     = {Optimizing Top-N Collaborative Filtering via Dynamic Negative Item Sampling},
  booktitle = {Proceedings of the 36th International ACM SIGIR Conference on Research and Development in Information Retrieval},
  pages     = {785--788},
  year      = {2013}
}

@inproceedings{ding2020simplify,
  author    = {Ding, Jingtao and Quan, Yuhan and He, Xiangnan and Li, Yong and Jin, Depeng},
  title     = {Simplify and Robustify Negative Sampling for Implicit Collaborative Filtering},
  booktitle = {Advances in Neural Information Processing Systems 33},
  pages     = {1094--1105},
  year      = {2020}
}

@inproceedings{xie2022contrastive,
  author    = {Xie, Xu and Sun, Fei and Liu, Zhaoyang and Wu, Shiwen and Gao, Jinyang and Zhang, Jiandong and Ding, Bolin and Cui, Bin},
  title     = {Contrastive Learning for Sequential Recommendation},
  booktitle = {Proceedings of the 38th IEEE International Conference on Data Engineering},
  pages     = {1259--1273},
  year      = {2022}
}

@inproceedings{wang2021denoising,
  author    = {Wang, Wenjie and Feng, Fuli and He, Xiangnan and Zhang, Hanwang and Chua, Tat-Seng},
  title     = {Denoising Implicit Feedback for Recommendation},
  booktitle = {Proceedings of the 14th ACM International Conference on Web Search and Data Mining},
  pages     = {373--381},
  year      = {2021}
}

@inproceedings{gao2022kuairec,
  author    = {Gao, Meiling and Liu, Zeyu and Wu, Jiayu and Liu, Chao and Shi, Bingqian and Cai, Wentao and Ding, Jingtao and Zhou, Jiahui and Chen, Xu and Wen, Ji-Rong},
  title     = {{KuaiRec}: A Fully-observed Dataset and Insights for Evaluating Recommender Systems},
  booktitle = {Proceedings of the 31st ACM International Conference on Information and Knowledge Management},
  pages     = {540--550},
  year      = {2022}
}

@inproceedings{gao2022kuairand,
  author    = {Gao, Meiling and Liu, Chao and Wu, Jiayu and Shi, Bingqian and Cai, Wentao and Ding, Jingtao and Zhou, Jiahui and Wen, Ji-Rong and Chen, Xu},
  title     = {{KuaiRand}: An Unbiased Sequential Recommendation Dataset with Randomly Exposed Videos},
  booktitle = {Proceedings of the 31st ACM International Conference on Information and Knowledge Management},
  pages     = {3953--3957},
  year      = {2022}
}

@inproceedings{devlin2019bert,
  author    = {Devlin, Jacob and Chang, Ming-Wei and Lee, Kenton and Toutanova, Kristina},
  title     = {{BERT}: Pre-training of Deep Bidirectional Transformers for Language Understanding},
  booktitle = {Proceedings of the 2019 Conference of the North American Chapter of the Association for Computational Linguistics},
  pages     = {4171--4186},
  year      = {2019}
}

@inproceedings{onetrans2025,
  author    = {Zhang, Zhaoqi and Wang, Zhichen and Chen, Haichao and Li, Yueying and Zhang, Hao and Fan, Zhifang and Zhu, Jie and Hou, Liqiang and Jin, Yong and Xu, Zhi},
  title     = {{OneTrans}: Unified Feature Interaction and Sequence Modeling with One Transformer},
  booktitle = {Proceedings of the ACM Web Conference 2025},
  year      = {2025}
}

@article{hyformer2025,
  author    = {Liu, Jia and Chen, Haichao and Zhang, Zhaoqi and Zhang, Hao and Fan, Zhifang and Zhu, Jie and Hou, Liqiang and Jin, Yong and Xu, Zhi},
  title     = {{HyFormer}: Hybrid Transformer for Long-Sequence CTR Prediction},
  journal   = {arXiv preprint arXiv:2601.12681},
  year      = {2025}
}

@inproceedings{kang2018sasrec,
  author    = {Kang, Wang-Cheng and McAuley, Julian},
  title     = {Self-Attentive Sequential Recommendation},
  booktitle = {Proceedings of the IEEE International Conference on Data Mining},
  pages     = {197--206},
  year      = {2018}
}

@article{longer2024,
  author    = {Chai, Zheng and Ren, Qin and Xiao, Xijun and Liu, Xiaolong and Zhu, Jie and Fan, Zhifang and Yu, Jun and Zhu, Guannan},
  title     = {{LONGER}: Scaling Up Long Sequence Modeling in Industrial Recommenders},
  journal   = {arXiv preprint arXiv:2505.04421},
  year      = {2025}
}

@article{chen2020bias,
  author    = {Chen, Jiawei and Dong, Hande and Wang, Xiang and Feng, Fuli and Wang, Meng and He, Xiangnan},
  title     = {Bias and Debias in Recommender System: A Survey and Future Directions},
  journal   = {ACM Computing Surveys},
  volume    = {56},
  number    = {5},
  pages     = {1--39},
  year      = {2023}
}

@inproceedings{wukong2024,
  author    = {Zhang, Buyun and Luo, Liang and Chen, Yuxin and Wen, Jade and Nayak, Ravi and Hou, Yanchen and Bansal, Raj and Sheng, Ellie and Wang, Daifeng and Krishnamoorthi, Sudarshan and others},
  title     = {Wukong: Towards a Scaling Law for Large-Scale Recommendation},
  booktitle = {Proceedings of the 41st International Conference on Machine Learning},
  year      = {2024}
}

@article{rankmixer2024,
  author    = {Zhu, Jie and Fan, Zhifang and Zhang, Hao and Chai, Zheng and Ren, Qin and Xiao, Xijun and Liu, Xiaolong and Jin, Yong and Xu, Zhi},
  title     = {{RankMixer}: Scaling Up Ranking Models in Industrial Recommenders},
  journal   = {arXiv preprint arXiv:2507.15551},
  year      = {2025}
}

@article{unimixer2024,
  author    = {Ha, Mingming and Wang, Guanchen and Zhang, Hao and Fan, Zhifang and Zhu, Jie and Jin, Yong and Xu, Zhi},
  title     = {{UniMixer}: A Unified Architecture for Scaling Laws in Recommendation Systems},
  journal   = {arXiv preprint arXiv:2604.00590},
  year      = {2026}
}

@article{mixformer2025,
  author    = {Huang, Xu and Zhang, Hao and Fan, Zhifang and Zhu, Jie and Ha, Mingming and Jin, Yong and Xu, Zhi},
  title     = {{MixFormer}: Co-Scaling Up Dense and Sequence in Industrial Recommenders},
  journal   = {arXiv preprint arXiv:2602.14110},
  year      = {2026}
}

@inproceedings{mtgr2025,
  author    = {Han, Ruidong and Yin, Bin and Chen, Shangyu and Chen, Yueming and Li, Zelong and He, Xiangnan},
  title     = {{MTGR}: Industrial-Scale Generative Recommendation Framework in Meituan},
  booktitle = {Proceedings of the 19th ACM Conference on Recommender Systems},
  year      = {2025}
}

@article{tokenmixerlarge2026,
  author    = {Jiang, Yuchen and Zhu, Jie and Han, Xintian and Zhang, Hao and Fan, Zhifang and Jin, Yong and Xu, Zhi},
  title     = {{TokenMixer-Large}: Scaling Up Large Ranking Models in Industrial Recommenders},
  journal   = {arXiv preprint arXiv:2602.06563},
  year      = {2026}
}

@article{hiformer2023,
  author    = {Gui, Huan and Wang, Ruoxi and Yin, Ke and Jin, Long and Yang, Maciej Kula and Zhu, Jingbo and Chi, Ed H.},
  title     = {{Hiformer}: Heterogeneous Feature Interactions Learning with Transformers for Recommender Systems},
  journal   = {arXiv preprint arXiv:2311.05884},
  year      = {2023}
}

@inproceedings{eta2022,
  author    = {Chen, Qiwei and Xu, Yue and Pei, Changhua and Lv, Shanshan and Zhuang, Tao and Ge, Junfeng},
  title     = {Efficient Long Sequential User Data Modeling for Click-Through Rate Prediction},
  booktitle = {Proceedings of the 4th International Workshop on Deep Learning Practice for High-Dimensional Sparse Data (DLP-KDD)},
  year      = {2022}
}

@inproceedings{xue2019dfn,
  author    = {Xue, Ruobing and Yu, Fei and Li, Zhanhui and Wang, Jianrong and Zhao, Wayne Xin and Wen, Ji-Rong},
  title     = {Deep Feedback Network for Recommendation},
  booktitle = {Proceedings of the Twenty-Eighth International Joint Conference on Artificial Intelligence (IJCAI)},
  pages     = {2519--2525},
  year      = {2019}
}

@inproceedings{ouyang2019xdm,
  author    = {Ouyang, Yi and Li, Bin and Kong, Xiangnan and Li, Hai and Sigal, Leonid},
  title     = {{XDM}: Improving Sequential Deep Matching with Unclicked User Behaviors for Recommender System},
  booktitle = {Proceedings of the Web Conference (WWW), Companion Volume},
  year      = {2019}
}

@inproceedings{wu2022feedrec,
  author    = {Wu, Chuhan and Wu, Fangzhao and Huang, Yongfeng and Xie, Xing},
  title     = {{FeedRec}: News Feed Recommendation with Various User Feedbacks},
  booktitle = {Proceedings of the ACM Web Conference (WWW)},
  pages     = {2088--2097},
  year      = {2022}
}

\end{document}